\documentclass[twocolumn,times]{aastex63}
\usepackage{graphicx}
\usepackage{makecell}
\usepackage{amsmath}
\usepackage{amsthm}
\usepackage{booktabs}
\usepackage{lineno}
\usepackage{float} 
\usepackage{threeparttable}
\usepackage{CJKutf8}
\usepackage[squaren]{SIunits}
\usepackage{xspace}
\usepackage{comment}
\usepackage{xcolor}
\usepackage{multirow}

\usepackage{array}

\setwatermarkfontsize{1.5in}

\shortauthors{Xiao ET AL.}
\begin{document}
\begin{CJK}{UTF8}{gbsn}

\title{Individual and Averaged Power Density Spectra of X-ray bursts from SGR J1935+2154: Quasiperiodic Oscillation Search and Slopes}
\author{Shuo Xiao*}
\affil{Guizhou Provincial Key Laboratory of Radio Astronomy and Data Processing, Guizhou Normal University, Guiyang 550001, People’s Republic of China}
\affil{School of Physics and Electronic Science, Guizhou Normal University, Guiyang 550001, People’s Republic of China;\\ xiaoshuo@gznu.edu.cn}

\author{Xiao-Bo Li*}
\affil{Key Laboratory of Particle Astrophysics, Institute of High Energy Physics, Chinese Academy of Sciences, Beijing 100049, China;\\  lixb@ihep.ac.cn}

\author{Wang-Chen Xue}
\affil{Key Laboratory of Particle Astrophysics, Institute of High Energy Physics, Chinese Academy of Sciences, Beijing 100049, China;\\  lixb@ihep.ac.cn}
\affil{University of Chinese Academy of Sciences, Chinese Academy of Sciences, Beijing 100049, China}

\author{Shao-Lin Xiong}
\affil{Key Laboratory of Particle Astrophysics, Institute of High Energy Physics, Chinese Academy of Sciences, Beijing 100049, China;\\  lixb@ihep.ac.cn}

\author{Shuang-Nan Zhang}
\affil{Key Laboratory of Particle Astrophysics, Institute of High Energy Physics, Chinese Academy of Sciences, Beijing 100049, China;\\  lixb@ihep.ac.cn}
\affil{University of Chinese Academy of Sciences, Chinese Academy of Sciences, Beijing 100049, China}

\author{Wen-Xi Peng}
\affil{Key Laboratory of Particle Astrophysics, Institute of High Energy Physics, Chinese Academy of Sciences, Beijing 100049, China;\\  lixb@ihep.ac.cn}

\author{Ai-Jun Dong}
\affil{Guizhou Provincial Key Laboratory of Radio Astronomy and Data Processing, Guizhou Normal University, Guiyang 550001, People’s Republic of China}
\affil{School of Physics and Electronic Science, Guizhou Normal University, Guiyang 550001, People’s Republic of China;\\ xiaoshuo@gznu.edu.cn}

\author{You-Li Tuo}
\affil{Institut für Astronomie und Astrophysik, University of Tübingen, Sand 1, 72076 Tübingen, Germany}

\author{Ce Cai}
\affil{College of Physics, Hebei Normal University, 20 South Erhuan Road, Shijiazhuang, 050024, China}

\author{Xi-Hong Luo}
\affil{Department of Physics, Chongqing University, Chongqing 400044, Peopleʼs Republic of China}

\author{Jiao-Jiao Yang}
\affil{Guizhou Provincial Key Laboratory of Radio Astronomy and Data Processing, Guizhou Normal University, Guiyang 550001, People’s Republic of China}
\affil{School of Physics and Electronic Science, Guizhou Normal University, Guiyang 550001, People’s Republic of China;\\ xiaoshuo@gznu.edu.cn}

\author{Yue Wang}
\affil{Key Laboratory of Particle Astrophysics, Institute of High Energy Physics, Chinese Academy of Sciences, Beijing 100049, China;\\  lixb@ihep.ac.cn}
\affil{University of Chinese Academy of Sciences, Chinese Academy of Sciences, Beijing 100049, China}

\author{Chao Zheng}
\affil{Key Laboratory of Particle Astrophysics, Institute of High Energy Physics, Chinese Academy of Sciences, Beijing 100049, China;\\  lixb@ihep.ac.cn}
\affil{University of Chinese Academy of Sciences, Chinese Academy of Sciences, Beijing 100049, China}

\author{Yan-Qiu Zhang}
\affil{Key Laboratory of Particle Astrophysics, Institute of High Energy Physics, Chinese Academy of Sciences, Beijing 100049, China;\\  lixb@ihep.ac.cn}
\affil{University of Chinese Academy of Sciences, Chinese Academy of Sciences, Beijing 100049, China}

\author{Jia-Cong Liu}
\affil{Key Laboratory of Particle Astrophysics, Institute of High Energy Physics, Chinese Academy of Sciences, Beijing 100049, China;\\  lixb@ihep.ac.cn}
\affil{University of Chinese Academy of Sciences, Chinese Academy of Sciences, Beijing 100049, China}

\author{Wen-Jun Tan}
\affil{Key Laboratory of Particle Astrophysics, Institute of High Energy Physics, Chinese Academy of Sciences, Beijing 100049, China;\\  lixb@ihep.ac.cn}
\affil{University of Chinese Academy of Sciences, Chinese Academy of Sciences, Beijing 100049, China}

\author{Chen-Wei Wang}
\affil{Key Laboratory of Particle Astrophysics, Institute of High Energy Physics, Chinese Academy of Sciences, Beijing 100049, China;\\  lixb@ihep.ac.cn}
\affil{University of Chinese Academy of Sciences, Chinese Academy of Sciences, Beijing 100049, China}

\author{Ping Wang}
\affil{Key Laboratory of Particle Astrophysics, Institute of High Energy Physics, Chinese Academy of Sciences, Beijing 100049, China;\\  lixb@ihep.ac.cn}

\author{Cheng-Kui Li}
\affil{Key Laboratory of Particle Astrophysics, Institute of High Energy Physics, Chinese Academy of Sciences, Beijing 100049, China;\\  lixb@ihep.ac.cn}

\author{Shu-Xu Yi}
\affil{Key Laboratory of Particle Astrophysics, Institute of High Energy Physics, Chinese Academy of Sciences, Beijing 100049, China;\\  lixb@ihep.ac.cn}

\author{Shi-Jun Dang}
\affil{Guizhou Provincial Key Laboratory of Radio Astronomy and Data Processing, Guizhou Normal University, Guiyang 550001, People’s Republic of China}
\affil{School of Physics and Electronic Science, Guizhou Normal University, Guiyang 550001, People’s Republic of China;\\ xiaoshuo@gznu.edu.cn}

\author{Lun-Hua Shang}
\affil{Guizhou Provincial Key Laboratory of Radio Astronomy and Data Processing, Guizhou Normal University, Guiyang 550001, People’s Republic of China}
\affil{School of Physics and Electronic Science, Guizhou Normal University, Guiyang 550001, People’s Republic of China;\\ xiaoshuo@gznu.edu.cn}

\author{Ru-shuang Zhao}
\affil{Guizhou Provincial Key Laboratory of Radio Astronomy and Data Processing, Guizhou Normal University, Guiyang 550001, People’s Republic of China}
\affil{School of Physics and Electronic Science, Guizhou Normal University, Guiyang 550001, People’s Republic of China;\\ xiaoshuo@gznu.edu.cn}

\author{Qing-Bo Ma}
\affil{Guizhou Provincial Key Laboratory of Radio Astronomy and Data Processing, Guizhou Normal University, Guiyang 550001, People’s Republic of China}
\affil{School of Physics and Electronic Science, Guizhou Normal University, Guiyang 550001, People’s Republic of China;\\ xiaoshuo@gznu.edu.cn}

\author{Wei Xie}
\affil{Guizhou Provincial Key Laboratory of Radio Astronomy and Data Processing, Guizhou Normal University, Guiyang 550001, People’s Republic of China}
\affil{School of Physics and Electronic Science, Guizhou Normal University, Guiyang 550001, People’s Republic of China;\\ xiaoshuo@gznu.edu.cn}

\author{Jian-Chao Feng}
\affil{Guizhou Provincial Key Laboratory of Radio Astronomy and Data Processing, Guizhou Normal University, Guiyang 550001, People’s Republic of China}
\affil{School of Physics and Electronic Science, Guizhou Normal University, Guiyang 550001, People’s Republic of China;\\ xiaoshuo@gznu.edu.cn}
\author{Bin Zhang}
\affil{Guizhou Provincial Key Laboratory of Radio Astronomy and Data Processing, Guizhou Normal University, Guiyang 550001, People’s Republic of China}
\affil{School of Physics and Electronic Science, Guizhou Normal University, Guiyang 550001, People’s Republic of China;\\ xiaoshuo@gznu.edu.cn}

\author{Zhen Zhang}
\affil{Key Laboratory of Particle Astrophysics, Institute of High Energy Physics, Chinese Academy of Sciences, Beijing 100049, China;\\  lixb@ihep.ac.cn}

\author{Ming-Yu Ge}
\affil{Key Laboratory of Particle Astrophysics, Institute of High Energy Physics, Chinese Academy of Sciences, Beijing 100049, China;\\  lixb@ihep.ac.cn}

\author{Shi-Jie Zheng}
\affil{Key Laboratory of Particle Astrophysics, Institute of High Energy Physics, Chinese Academy of Sciences, Beijing 100049, China;\\  lixb@ihep.ac.cn}

\author{Li-Ming Song}
\affil{Key Laboratory of Particle Astrophysics, Institute of High Energy Physics, Chinese Academy of Sciences, Beijing 100049, China;\\  lixb@ihep.ac.cn}

\author{Qi-Jun Zhi}
\affil{Guizhou Provincial Key Laboratory of Radio Astronomy and Data Processing, Guizhou Normal University, Guiyang 550001, People’s Republic of China}
\affil{School of Physics and Electronic Science, Guizhou Normal University, Guiyang 550001, People’s Republic of China;\\ xiaoshuo@gznu.edu.cn}




\begin{abstract}
The study of quasi-periodic oscillations (QPOs) and power density spectra (PDS) continuum properties can help shed light on the still illusive emission physics of magnetars and as a window into the interiors of neutron stars using asteroseismology. In this work, we employ a Bayesian method to search for the QPOs in the hundreds of X-ray bursts from SGR J1935+2154 observed by {\it Insight}-HXMT, GECAM and Fermi/GBM from July 2014 to January 2022. Although no definitive QPO signal (significance $>3\sigma$) is detected in individual bursts or the averaged periodogram of the bursts grouped by duration, we identify several bursts exhibiting possible QPO at $\sim$ 40 Hz, which is consistent with that reported in the X-ray burst associated with FRB 200428. We investigate the PDS continuum properties and find that the distribution of the PDS slope in the simple power-law model peaks $\sim$ 2.5, which is consistent with other magnetars but higher than 5/3 commonly seen in gamma-ray bursts. Besides, the distribution of the break frequency in the broken power-law model peaks at $\sim$ 60 Hz. Finally, we report that the power-law index of PDS has an anti-correlation and power-law dependence on the burst duration as well as the minimum variation timescale.

\end{abstract}


\section{Introduction}
Quasi-periodic oscillations (QPOs) have been reported in some magnetars, including SGR 1806-20 \citep{israel2005discovery, watts2006detection, 2006ApJ...653..593S, huppenkothen2014quasi}, SGR 1900+14 \citep{strohmayer2005discovery}, SGR J1550-5418 \citep{2014ApJ...793..129H}, SGR J1935+2154 \citep{li2022quasi}, an extragalactic magnetar (GRB 200415A) \citep{castro2021very} and even in a magnetar in a compact binary merger (GRB 211211A) \citep{xiao2022quasi}. \cite{roberts2023quasi} recently reported 42 Hz QPO candidates in the $vF_v$ spectrum peak energy ($E_{\rm p}$) of a burst from SGR J1935+2154, though statistically may not be very significant (e.g. \citealp{pumpe2018search}). The QPOs can be explained as originating from oscillations in the crustal movement of magnetars \citep{2014ApJ...787..128H,2019ApJ...871...95M}, therefore, in principle, QPO can be used to constrain both the equation of state and the interior magnetic field of the neutron star through the identification of QPO frequency via particular global seismic modes \citep{huppenkothen2013quasi}. These QPO studies have opened up the possibility of studying neutron star interiors via asteroseismology.

In our recent work, we performed a systematic analysis of the spectral lags \citep{xiao2023discovery} and minimum variation timescales (MVTs) \citep{xiao2023minimum} of the X-ray burst from SGR J1935+2154, and reported that the characteristics are significantly different from gamma-ray bursts (GRBs). Besides, the MVT of the burst associated with FRB 200428 is larger compared to most other bursts, which implies its special radiation mechanism \citep{xiao2023minimum}. Therefore, regarding the recently reported interesting QPO in the burst associated with FRB 200428 from SGR J1935+2154 \citep{li2022quasi}, we 
wonder if this magnetar is special compared to others. For example, are there QPOs in other X-ray bursts from this magnetar? Do their power density spectra (PDSs) have similar continuum properties   \citep{huppenkothen2013quasi} or QPO candidates like other magnetars in their average PDSs (e.g. \citealp{2014ApJ...793..129H})? However, to date these analyses have not been systematically studied, which is the focus of this work.

PDS is not only the most common method for QPOs search in frequency domain \citep{2010MNRAS.402..307V,huppenkothen2013quasi,guidorzi2016individual,xiao2022search}, the PDS continuum properties also contain valuable information,  \cite{huppenkothen2013quasi} and  \cite{guidorzi2016individual} indicated the differences between the distributions of the PDS slopes of gamma-ray
bursts (GRBs), active galactic nuclei (AGNs) and 21 magnetar bursts, \cite{guidorzi2016individual} then suggest the PDS
slope depends on the relative strength of the fast component with timescale $<\sim 1$ s \citep{2003PASJ...55..345S,vetere2006slow,2012ApJ...748..134G} in the light curve. For the very active SGR J1935+2154, there are about 700 bursts from it observed by {\it Insight}-HXMT, GECAM and Fermi/GBM from July 2014 to Jan 2022, which are beneficial for PDS analyses. Besides, these three satellites have dead times and time resolutions in the order of microseconds or sub-microsecond \citep{liu2020high,xiao2020deadtime,xiao2022ground,meegan2009fermi,liu2021sipm}, which allows high-frequency QPO searches.

In this work, we present our sample selection and analysis methods in Section 2 and give the results of the PDS slopes and QPO searches in Section 3. Finally, a discussion and summary are given in Section 4.

\section{SAMPLE SELECTION and Methodology}

\subsection{Data selection}
The bursts from SGR J1935+2154 are collected from our previous work \citep{xiao2023minimum}. After removing the bursts with saturation or “timing glitches”, a total of 40, 82 and 503 bursts from SGR J1935+2154 are collected for HXMT, GECAM, and GBM, respectively.

The energy range used in our analysis is 8-100 keV. For GECAM and GBM, those detectors with incident angles of SGR J1935+2154 less than 60 degrees are selected to improve the statistics. Note that the observed energy range of GECAM may in fact be greater than 10 keV due to instrumental effects.

The time range of the light curve in the QPO search for an individual burst is selected using the Bayesian block method, which is a nonparametric modeling technique to detect and characterize local variability in time series of sequential data and can find the optimal segmentation of the burst \citep{scargle2013studies}. The time bin size of the light curves is selected as 0.1 ms, which can achieve a Nyquist frequency of 5000 Hz.

\begin{figure}[http]
\centering
\includegraphics[width=\columnwidth]{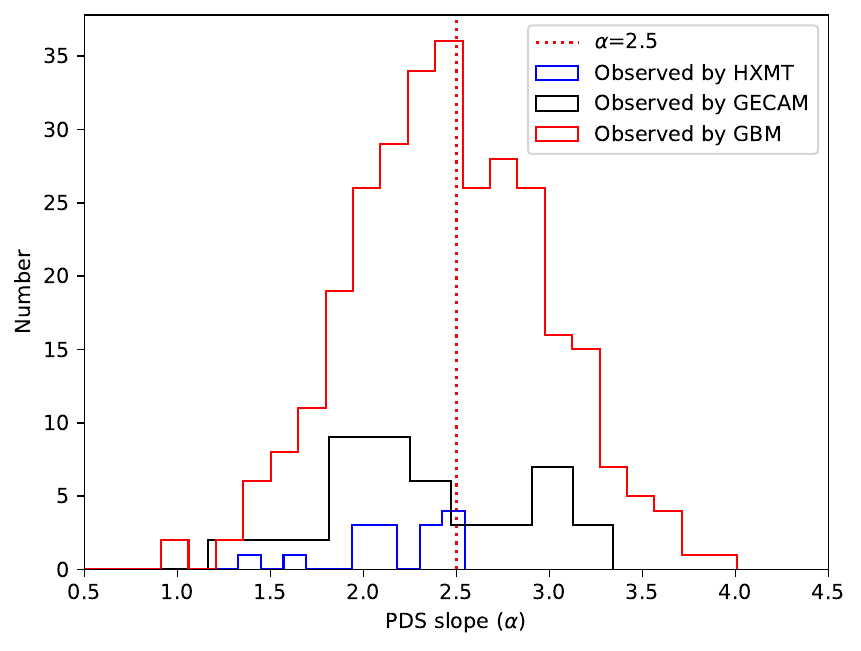}
\caption{Distribution of the PDS slopes (i.e. power-law index $\alpha$) in PLC model for bursts from SGR J1935+2154 observed by GBM, GECAM and HXMT, respectively, the red dotted line represents the peak 2.5 of the distribution.}\label{alphas}
\end{figure}

\begin{figure*}
\centering
\begin{minipage}[t]{0.45\textwidth}
\centering
\includegraphics[width=\columnwidth]{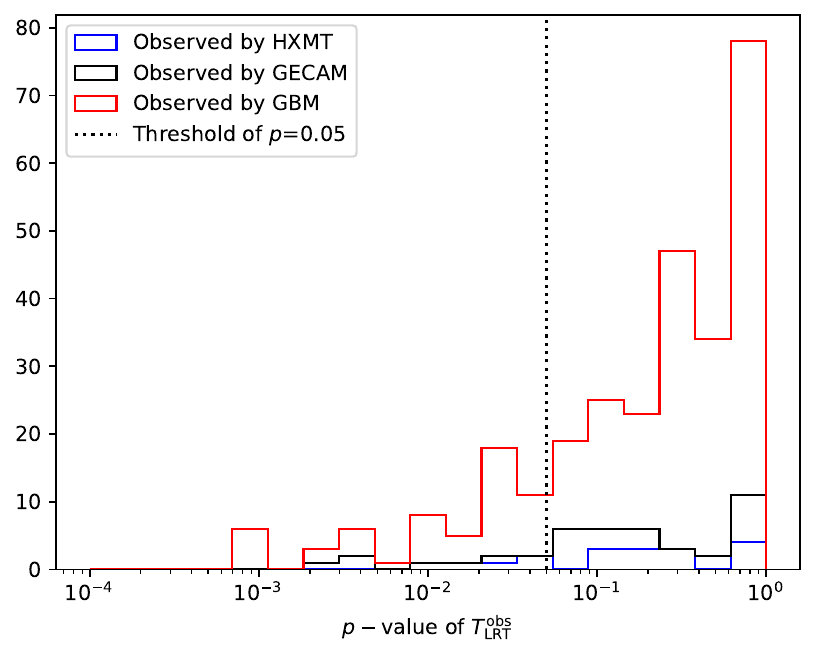}
\end{minipage}
\begin{minipage}[t]{0.47\textwidth}
\centering
\includegraphics[width=\columnwidth]{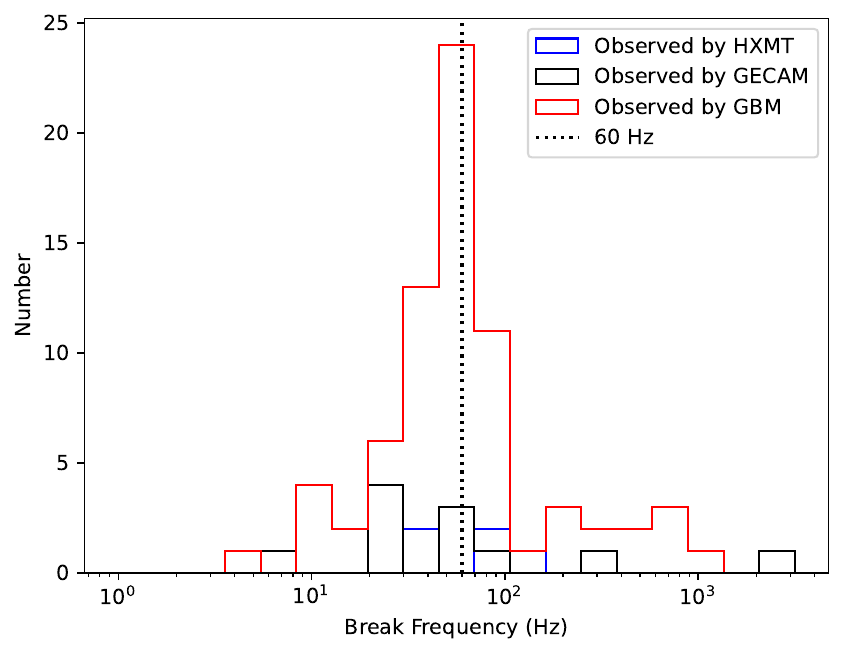}
\end{minipage}
\caption{Left panel: distributions of $p$-values corresponding to the likelihood ratios of the PLC and BPLC models, the black dotted line represents the 0.05 threshold we set, i.e., BPLC is considered the optimal model when $p$-value $<$0.05. Right panel: distributions of the break frequency (i.e. the break frequency between the low-frequency and high-frequency power law) for the optimal model is BPLC, the black dotted line represents the peak 60 Hz of the distribution. }\label{pl_bpl_p}
\end{figure*}

\begin{figure}[http]
\centering
\includegraphics[width=\columnwidth]{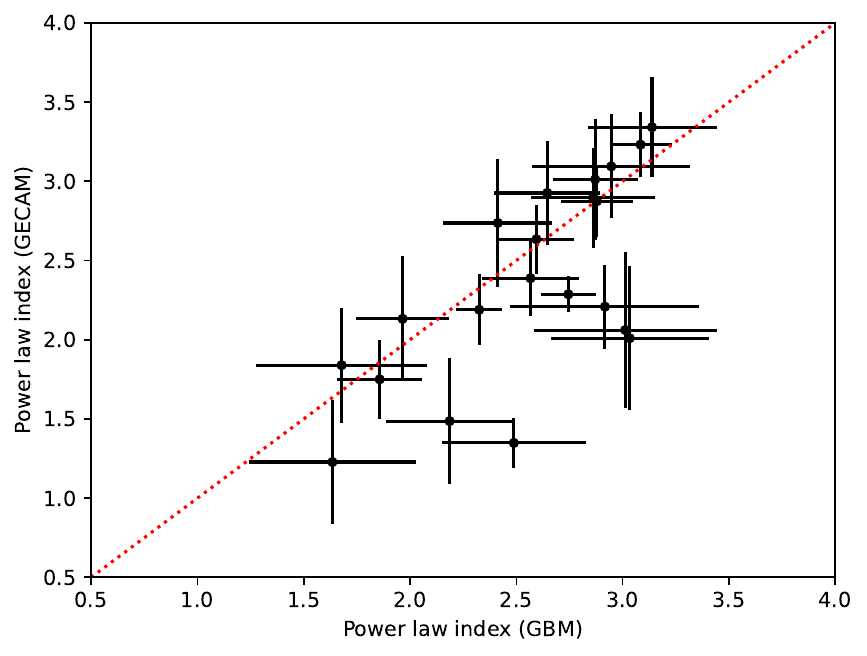}
\caption{Consistency check of the power-law index measured by GBM and GECAM. The red dashed lines represent the equality line.}\label{GBM_GECAM}
\end{figure}

\begin{figure*}
\centering
\begin{minipage}[t]{0.46\textwidth}
\centering
\includegraphics[width=\columnwidth]{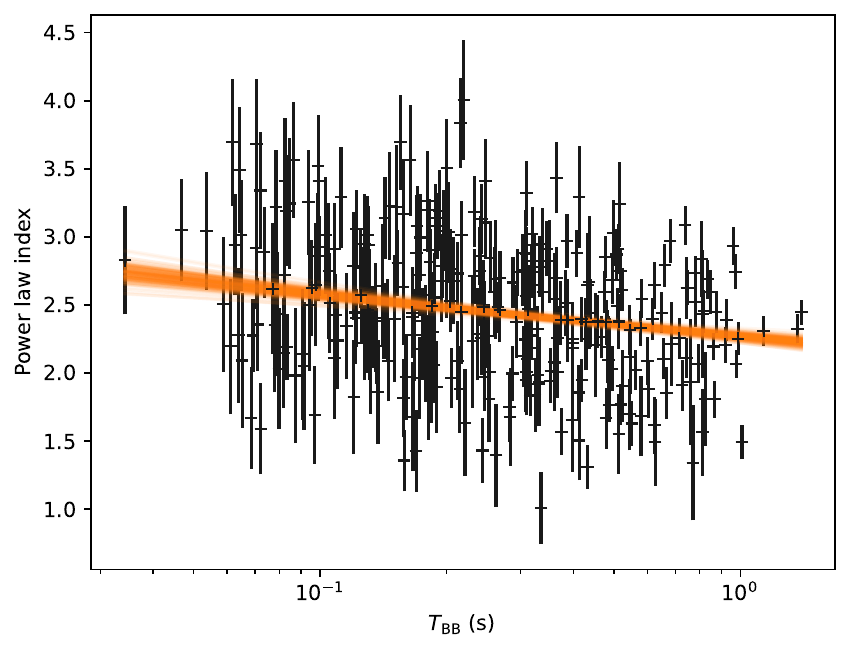}
\end{minipage}
\begin{minipage}[t]{0.46\textwidth}
\centering
\includegraphics[width=\columnwidth]{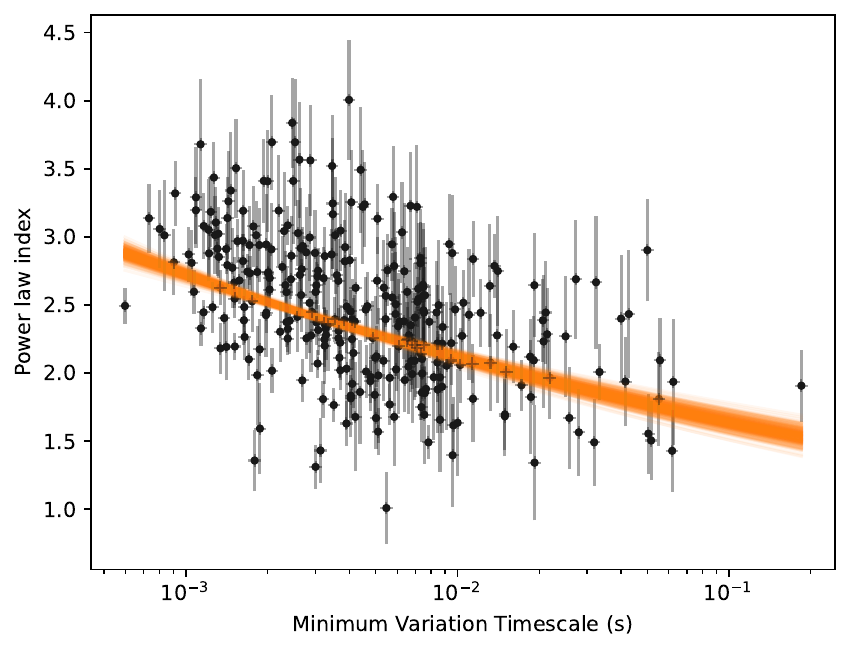}
\end{minipage}
\caption{Left panel: power-law index of PDS vs. burst duration ($T_{\rm BB}$), the fitted yellow line $\alpha=2.27_{-0.02}^{+0.02}\times T_{\rm BB}^{-0.06_{-0.01}^{+0.01}}$ is obtained by MCMC. Right panel: power-law index of PDS vs. minimum variation timescale, the fitted yellow line is $\alpha=1.28_{-0.06}^{+0.06}\times \Delta t_{\rm MVT}^{-0.11_{-0.01}^{+0.01}}$.}\label{alpha_T}
\end{figure*}

\begin{figure*}
\centering
\begin{minipage}[t]{0.46\textwidth}
\centering
\includegraphics[width=\columnwidth]{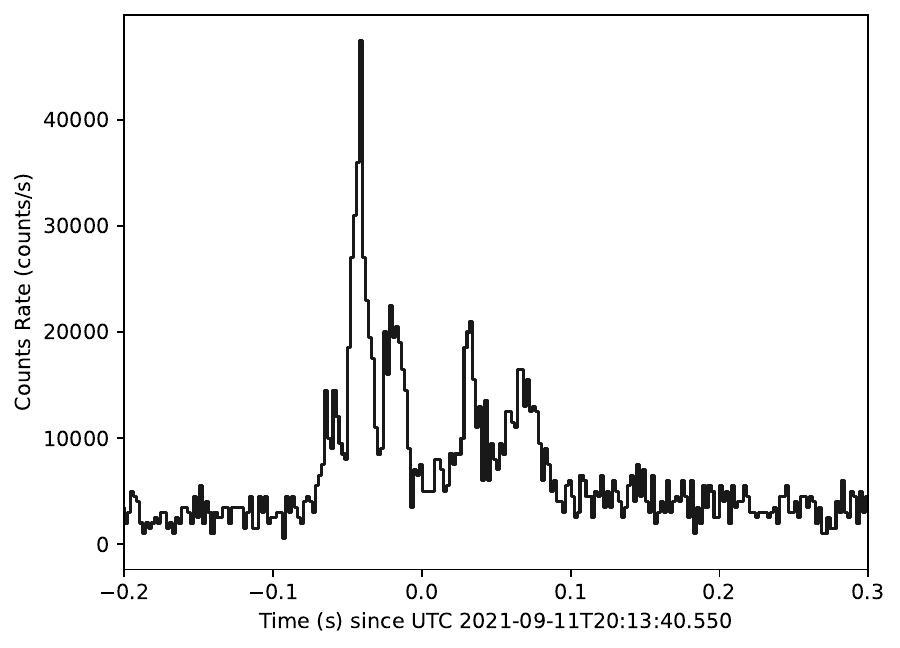}
\end{minipage}
\begin{minipage}[t]{0.45\textwidth}
\centering
\includegraphics[width=\columnwidth]{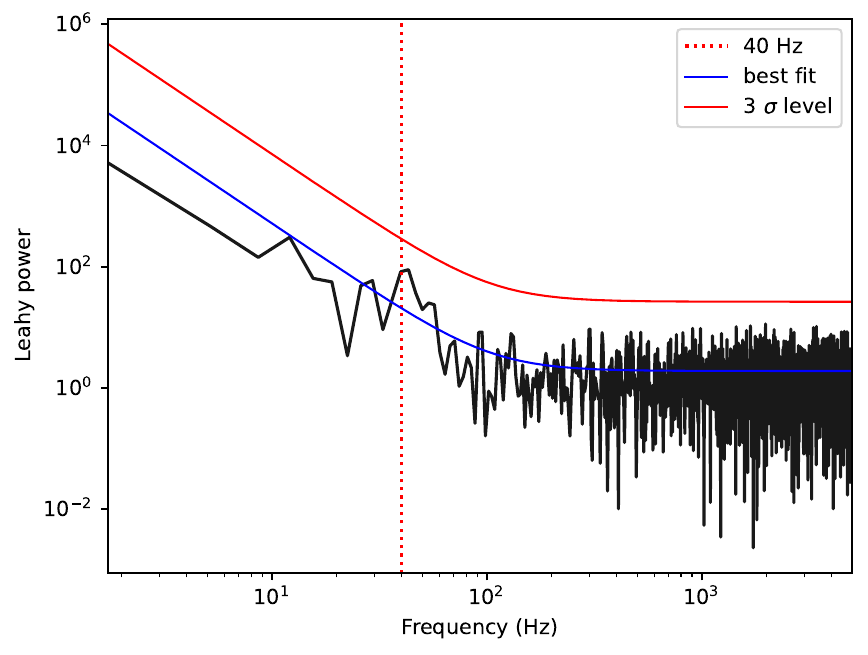}
\end{minipage}
\caption{Left panel: the 2 ms binned light curve of a burst (UTC 2021-09-11T20:13:40.550) from SGR J1935+2154 observed by GECAM. Right panel: unbinned PDS of the burst, the blue line is fitted by the PLC model, and the red line represents the 3 $\sigma$ level. The red dotted line represents the possible signal around 40 Hz.}\label{lcps0}
\end{figure*}

\begin{figure*}
\centering
\begin{minipage}[t]{0.46\textwidth}
\centering
\includegraphics[width=\columnwidth]{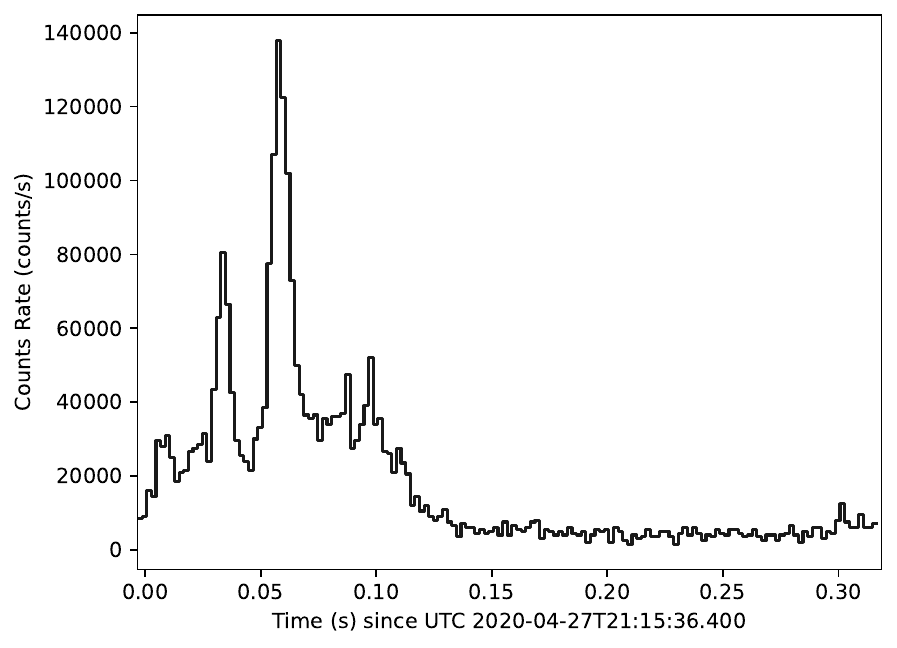}
\end{minipage}
\begin{minipage}[t]{0.45\textwidth}
\centering
\includegraphics[width=\columnwidth]{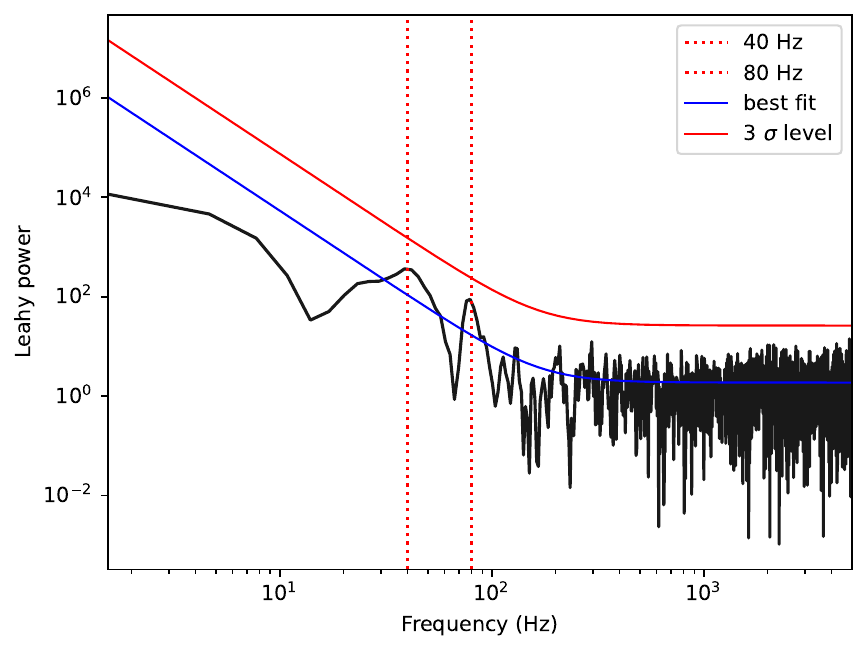}
\end{minipage}
\caption{Left panel: the 2 ms binned light curve of a burst (UTC 2020-04-27T21:15:36.400) from SGR J1935+2154 observed by the GBM for NaIs 3, 4, 6, 7 and 8. Right panel: unbinned PDS of the burst, the blue line is fitted by the PLC model, and the red line represents the 3 $\sigma$ level. The 40 Hz and 80 Hz are QPO candidates.}\label{lcps1}
\end{figure*}

\begin{figure*}
\centering
\begin{minipage}[t]{0.46\textwidth}
\centering
\includegraphics[width=\columnwidth]{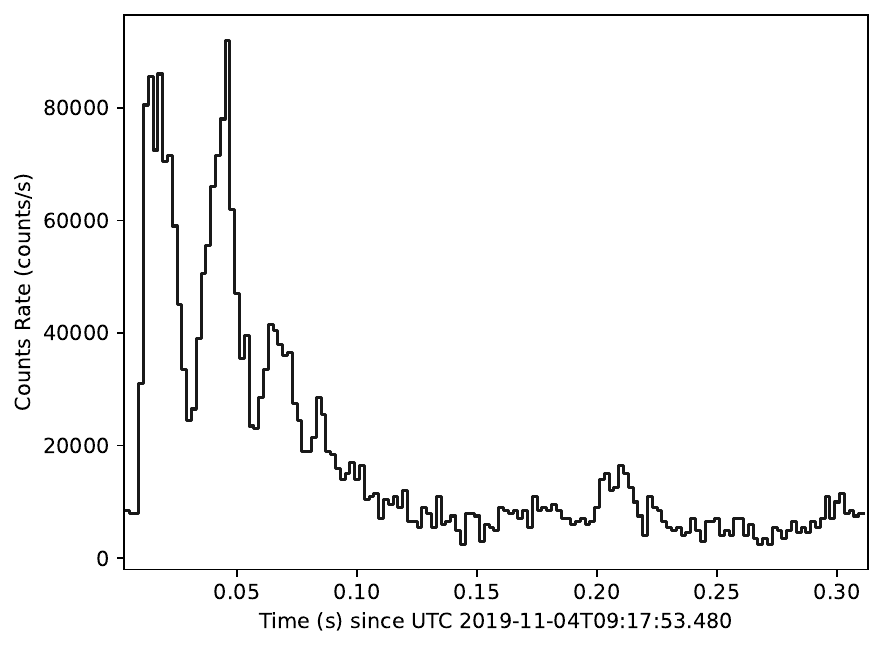}
\end{minipage}
\begin{minipage}[t]{0.45\textwidth}
\centering
\includegraphics[width=\columnwidth]{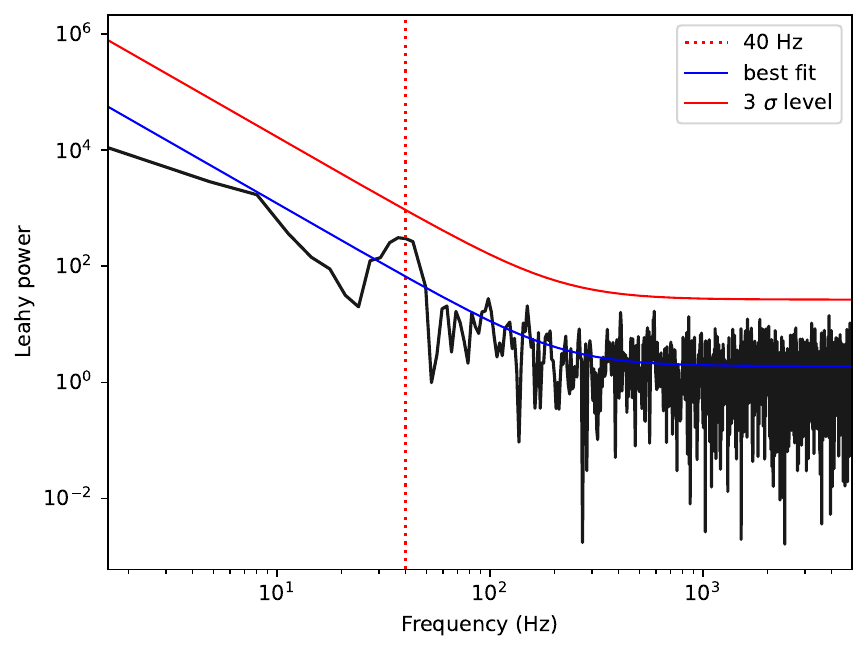}
\end{minipage}
\caption{Left panel: the 2 ms binned light curve of a burst (UTC 2019-11-04T09:17:53.480) from SGR J1935+2154 observed by GBM for NaIs 3, 4, 6, 7 and 8. Right panel: the blue line is fitted by the PLC model, and the red line represents the 3 $\sigma$ level. The 40 Hz is a QPO candidate.}\label{lcps2}
\end{figure*}

\begin{figure*}
\centering
\begin{minipage}[t]{0.46\textwidth}
\centering
\includegraphics[width=\columnwidth]{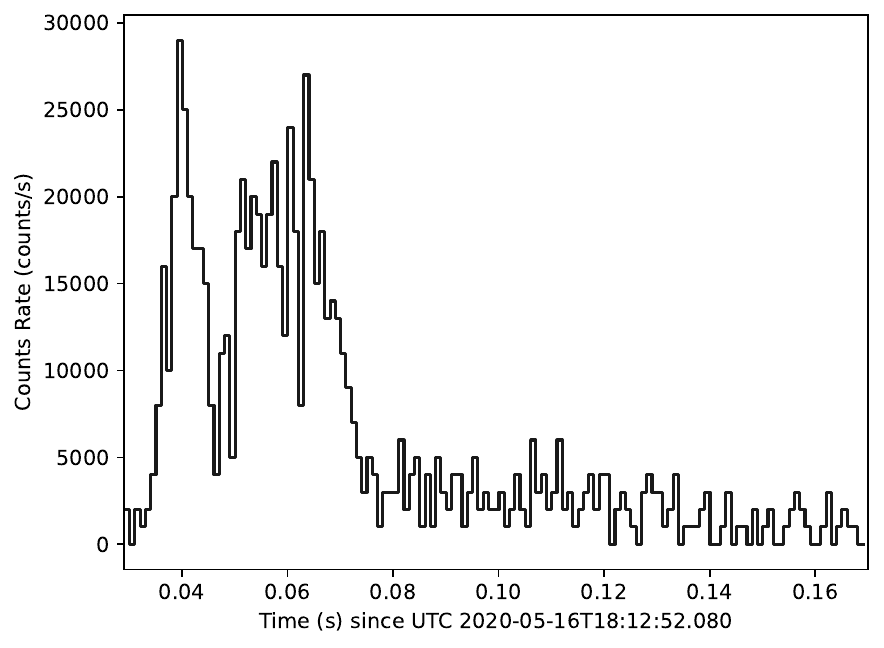}
\end{minipage}
\begin{minipage}[t]{0.45\textwidth}
\centering
\includegraphics[width=\columnwidth]{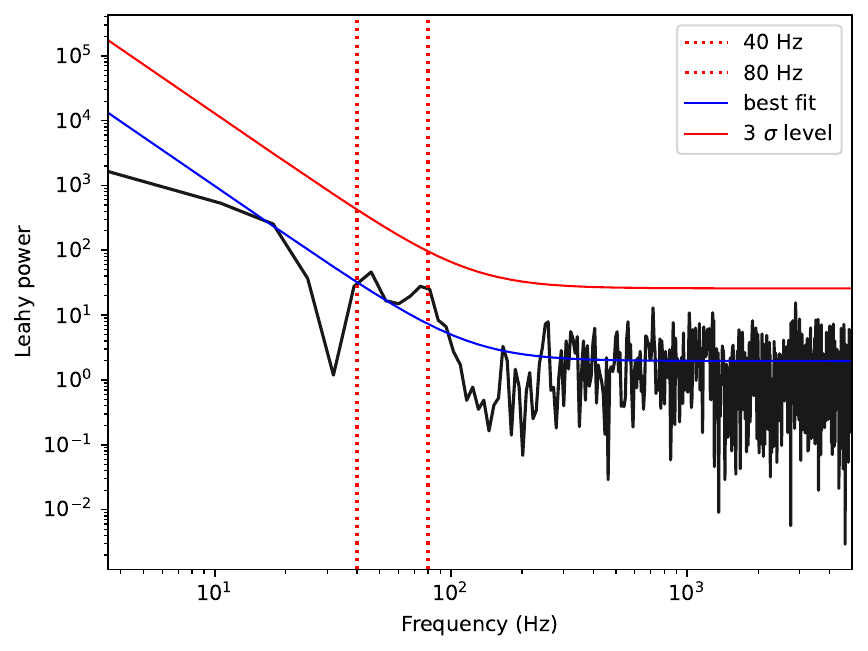}
\end{minipage}
\caption{Left panel: the 1 ms binned light curve of a burst (UTC 2020-05-16T18:12:52.080) from SGR J1935+2154 observed by HXMT. Right panel: the blue line is fitted by the PLC model, and the red line represents the 3 $\sigma$ level. Note that there are also QPO candidates around 40 Hz and 80 Hz.}\label{lcps3}
\end{figure*}

\begin{figure*}
\centering
\begin{minipage}[t]{0.49\textwidth}
\centering
\includegraphics[width=\columnwidth]{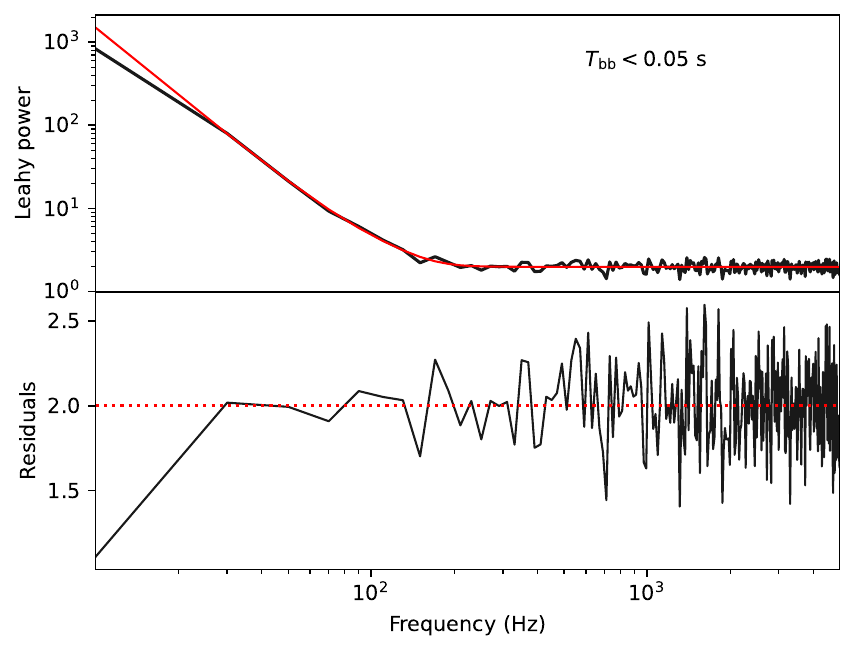}
\end{minipage}
\begin{minipage}[t]{0.49\textwidth}
\centering
\includegraphics[width=\columnwidth]{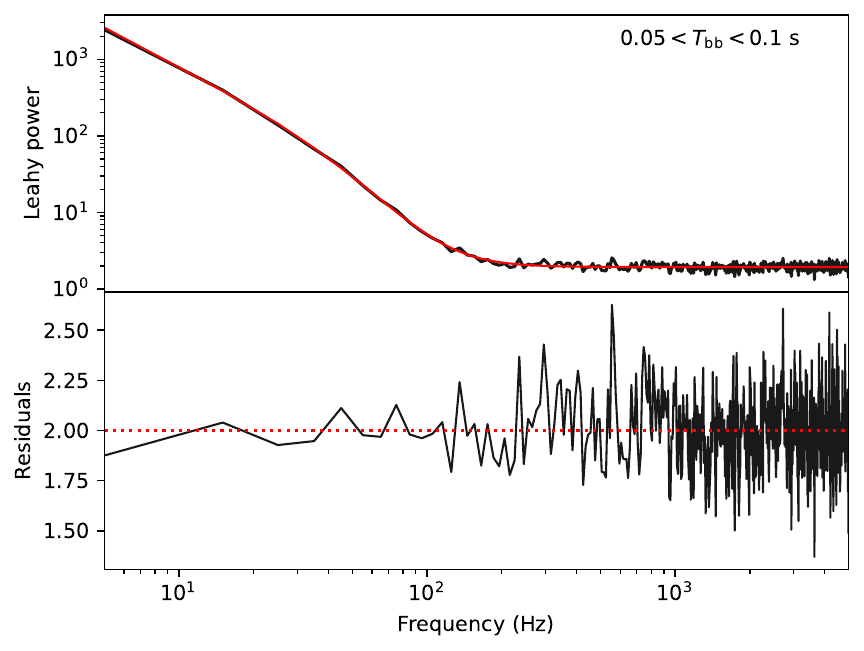}
\end{minipage}
\begin{minipage}[t]{0.49\textwidth}
\centering
\includegraphics[width=\columnwidth]{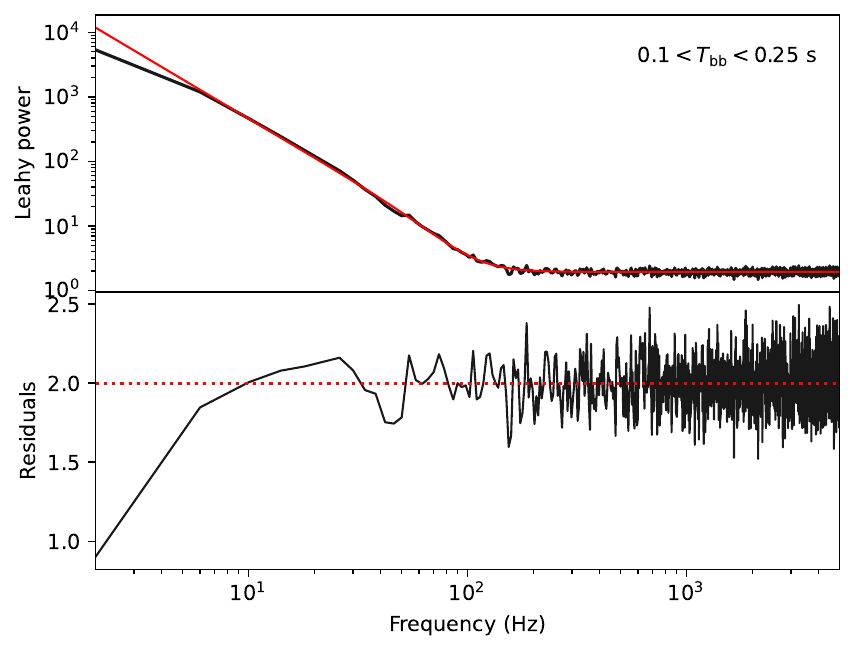}
\end{minipage}
\begin{minipage}[t]{0.49\textwidth}
\centering
\includegraphics[width=\columnwidth]{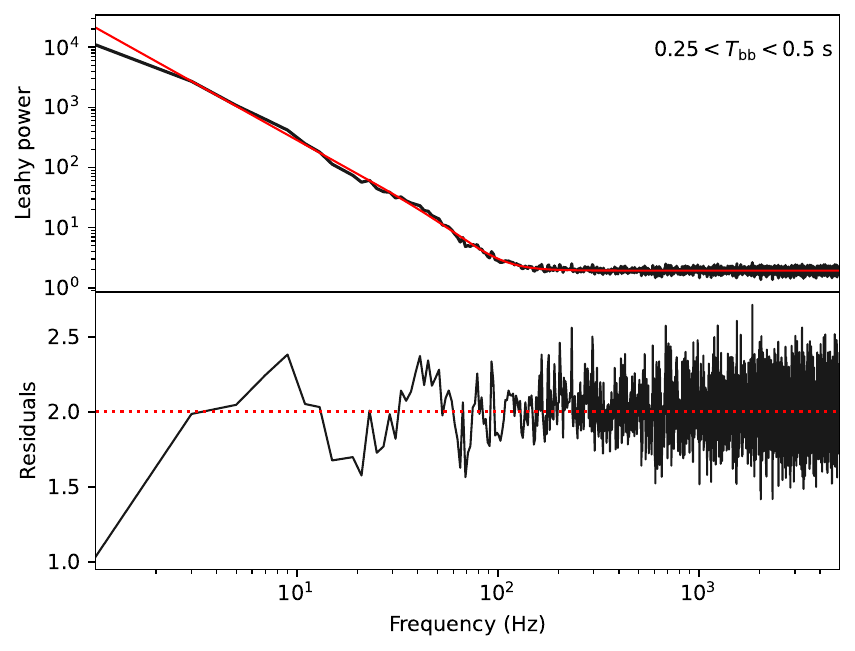}
\end{minipage}
\begin{minipage}[t]{0.49\textwidth}
\centering
\includegraphics[width=\columnwidth]{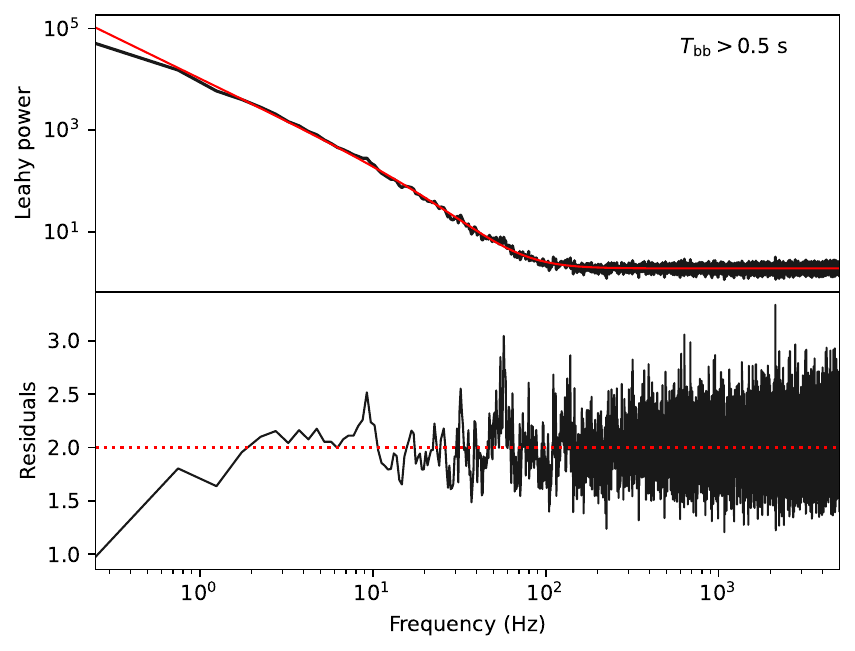}
\end{minipage}
\begin{minipage}[t]{0.49\textwidth}
\centering
\includegraphics[width=\columnwidth]{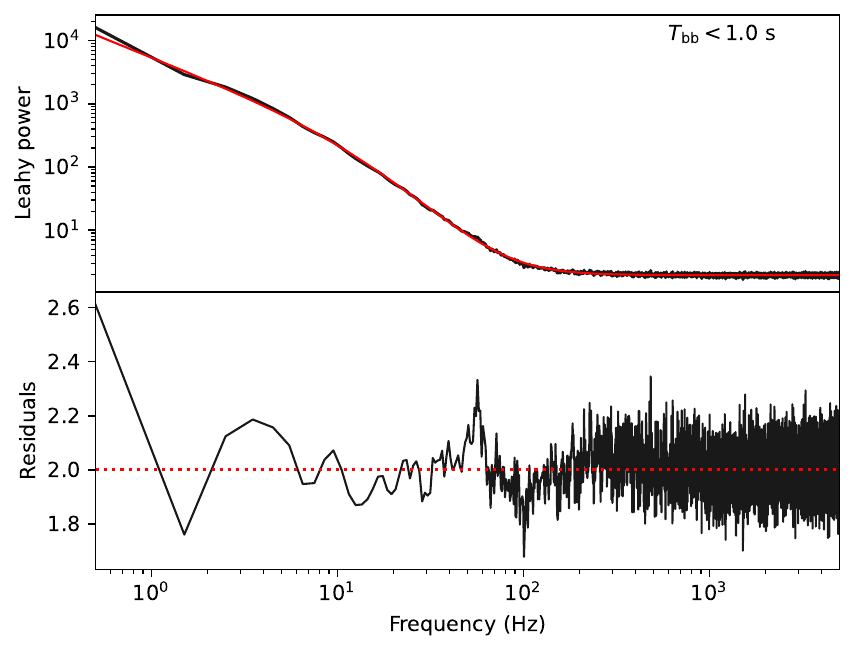}
\end{minipage}
\caption{Averaged periodograms of the bursts with different durations observed by GBM, the red lines are obtained by fitting with BPLC model. The bottom right panel shows the average power spectrum obtained by choosing 1-second duration for all bursts.}\label{ave_dua}
\end{figure*}

\begin{figure*}
\centering
\begin{minipage}[t]{0.49\textwidth}
\centering
\includegraphics[width=\columnwidth]{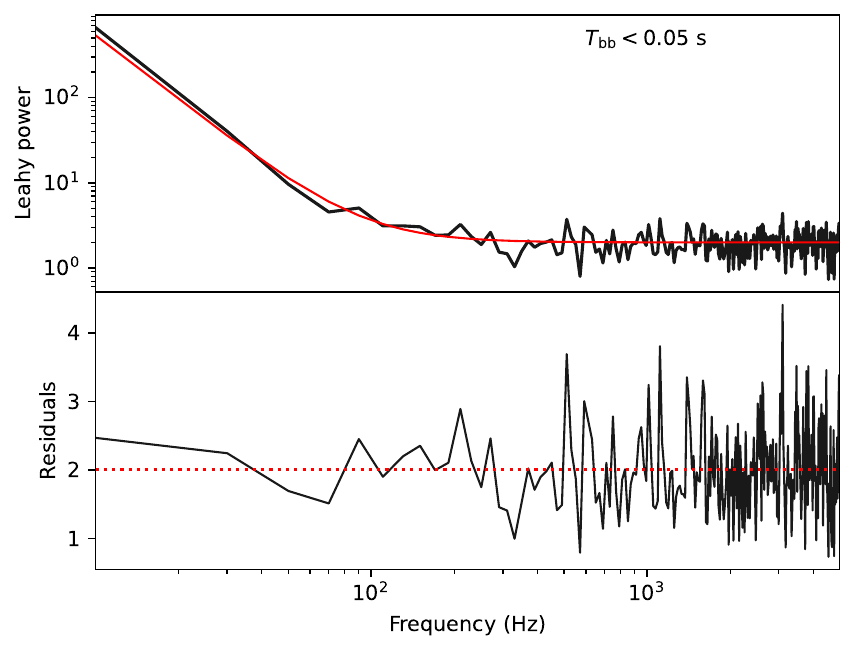}
\end{minipage}
\begin{minipage}[t]{0.49\textwidth}
\centering
\includegraphics[width=\columnwidth]{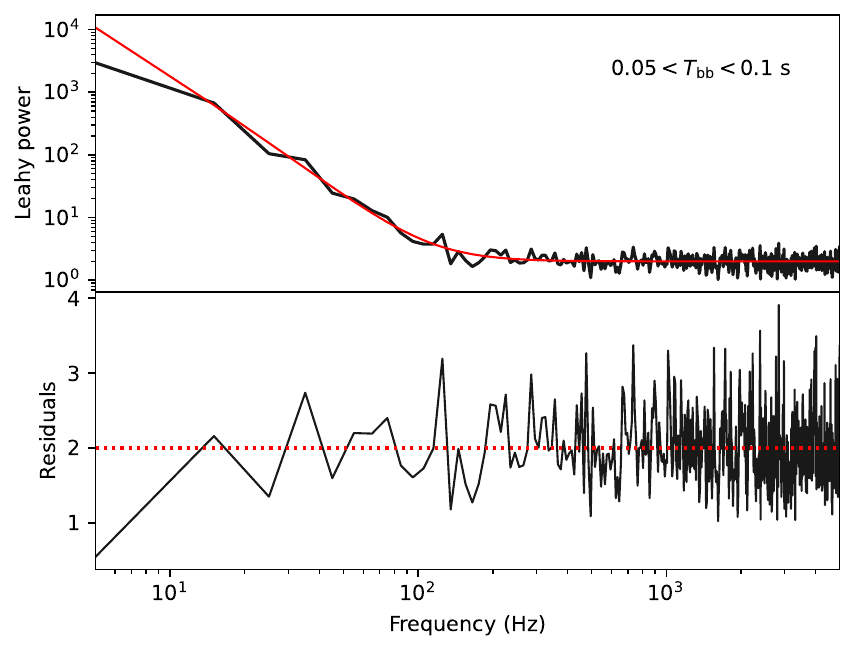}
\end{minipage}
\begin{minipage}[t]{0.49\textwidth}
\centering
\includegraphics[width=\columnwidth]{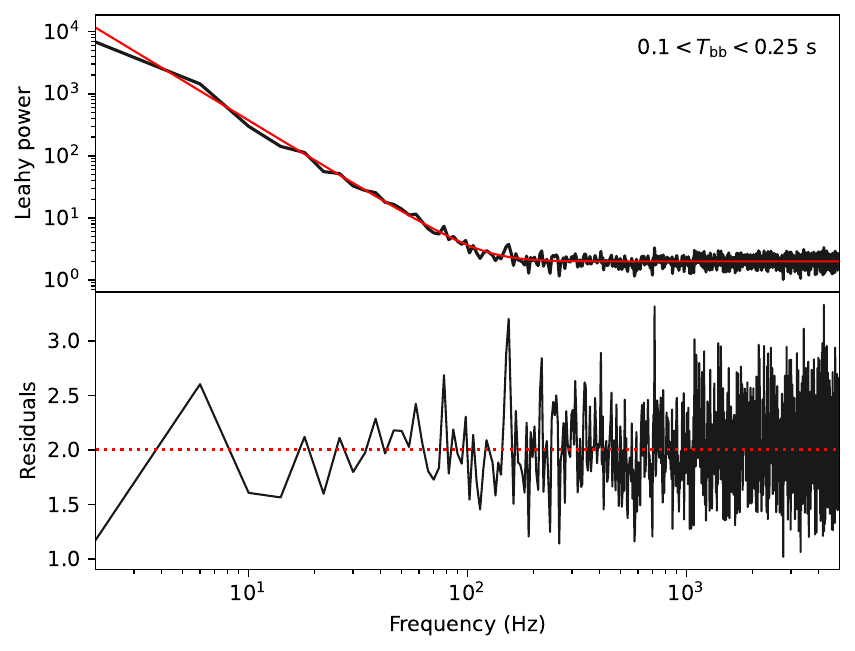}
\end{minipage}
\begin{minipage}[t]{0.49\textwidth}
\centering
\includegraphics[width=\columnwidth]{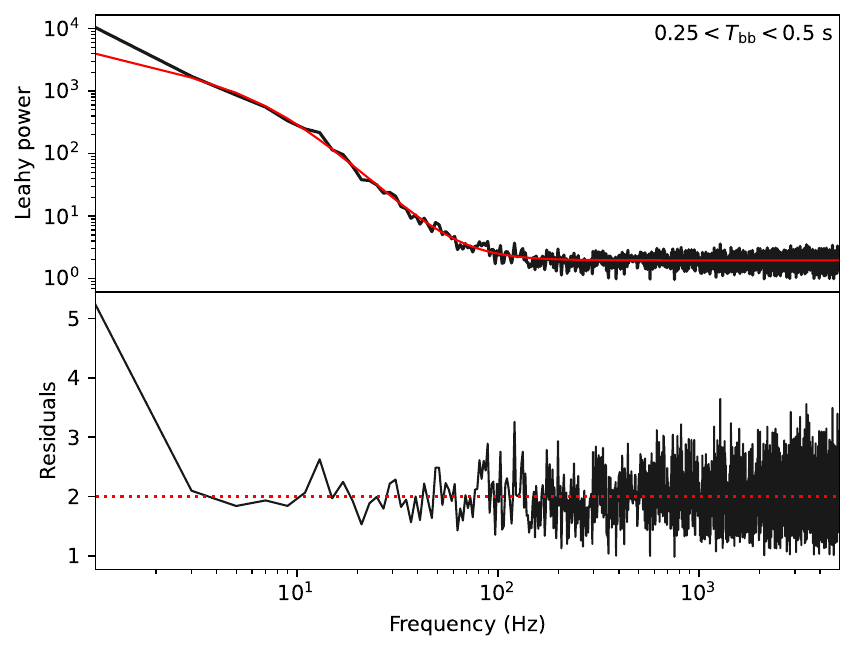}
\end{minipage}
\begin{minipage}[t]{0.49\textwidth}
\centering
\includegraphics[width=\columnwidth]{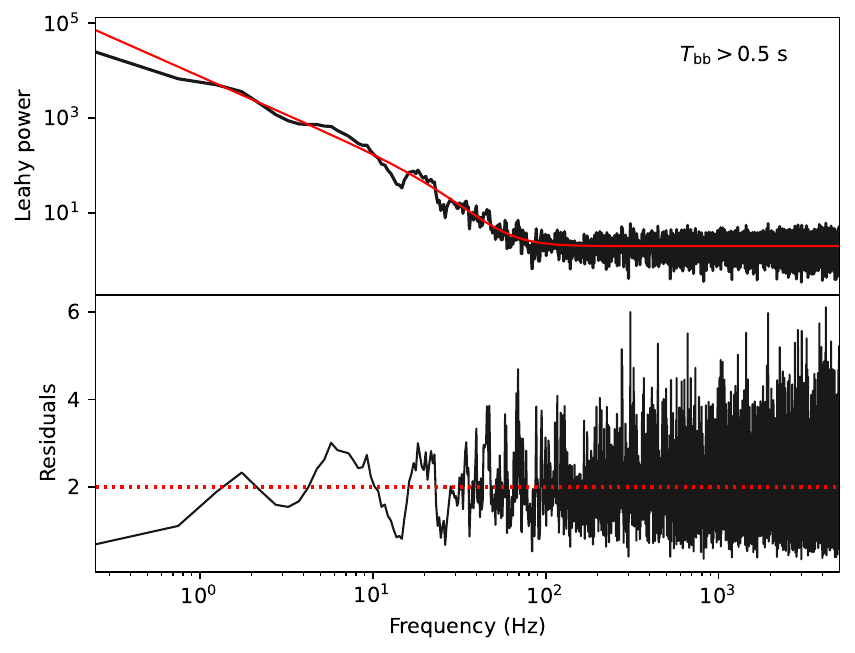}
\end{minipage}
\begin{minipage}[t]{0.49\textwidth}
\centering
\includegraphics[width=\columnwidth]{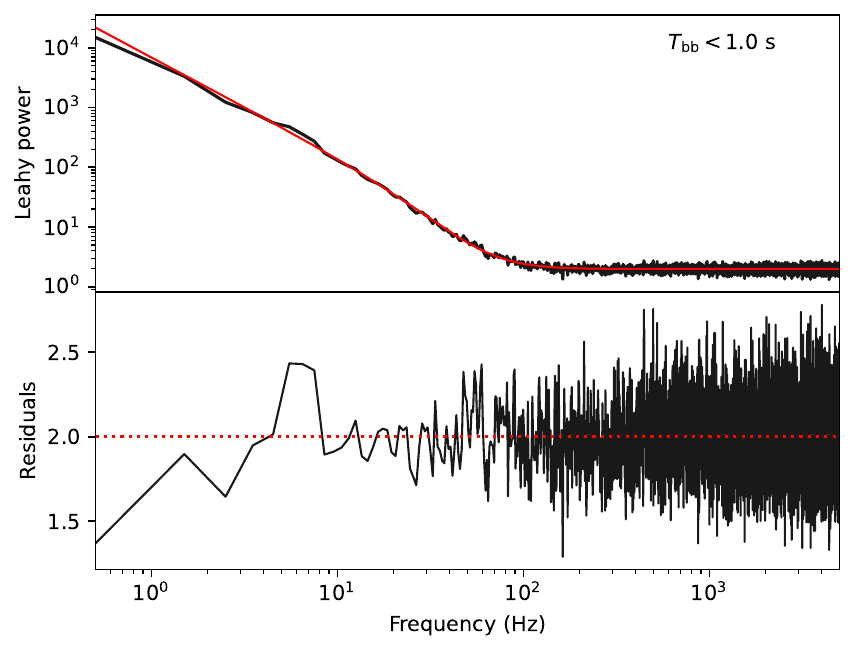}
\end{minipage}
\caption{Averaged periodograms of the bursts with different durations observed by GECAM, the red lines are obtained by fitting with BPLC model. The bottom right panel shows the average power spectrum obtained by choosing 1-second duration for all bursts.}\label{ave_dua_gecam}
\end{figure*}

\begin{figure*}
\centering
\begin{minipage}[t]{0.49\textwidth}
\centering
\includegraphics[width=\columnwidth]{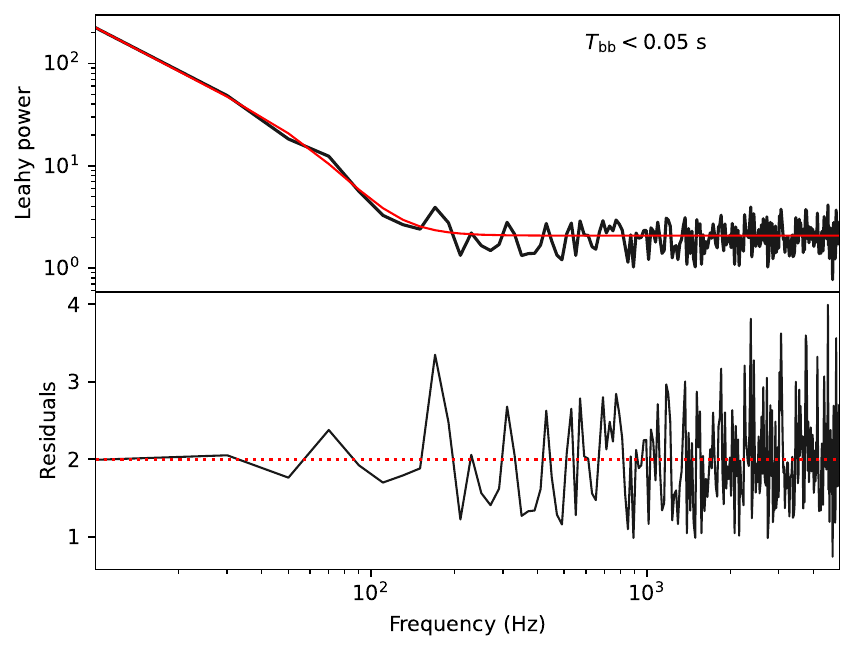}
\end{minipage}
\begin{minipage}[t]{0.49\textwidth}
\centering
\includegraphics[width=\columnwidth]{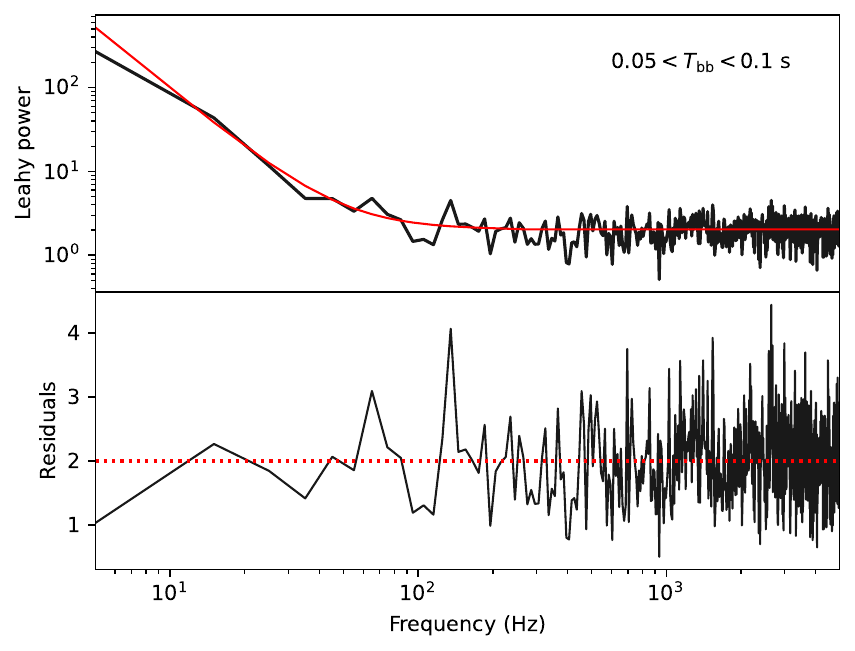}
\end{minipage}
\begin{minipage}[t]{0.49\textwidth}
\centering
\includegraphics[width=\columnwidth]{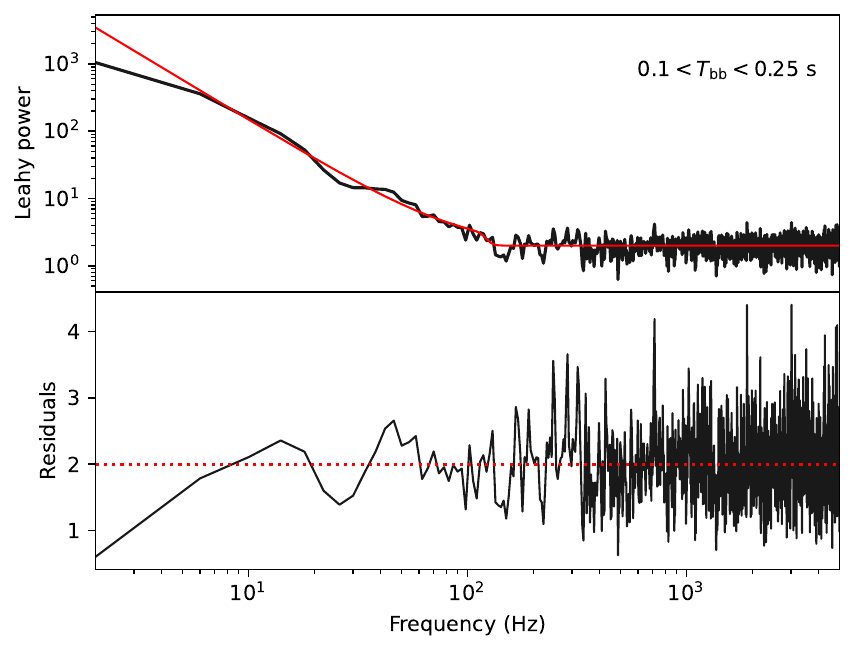}
\end{minipage}
\begin{minipage}[t]{0.49\textwidth}
\centering
\includegraphics[width=\columnwidth]{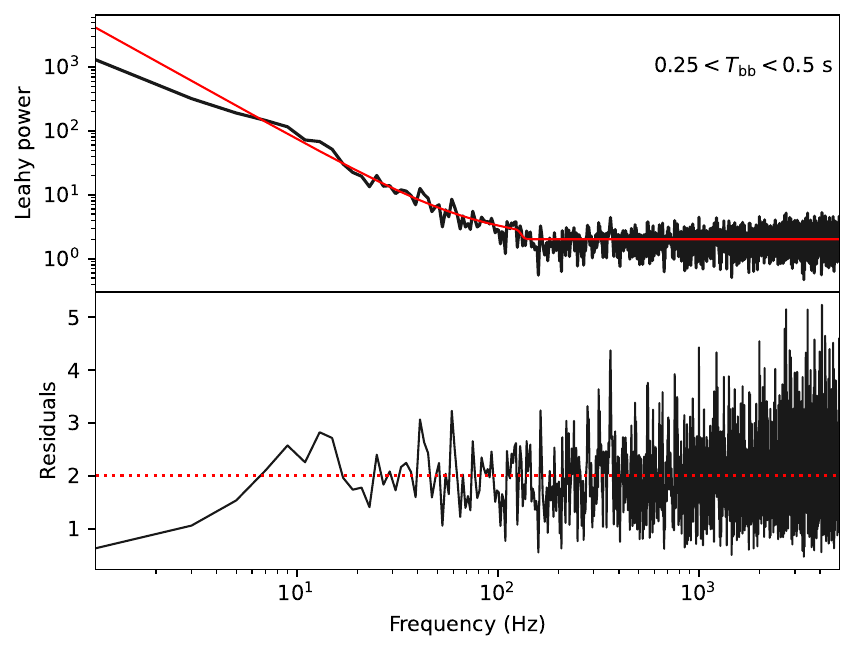}
\end{minipage}
\begin{minipage}[t]{0.49\textwidth}
\centering
\includegraphics[width=\columnwidth]{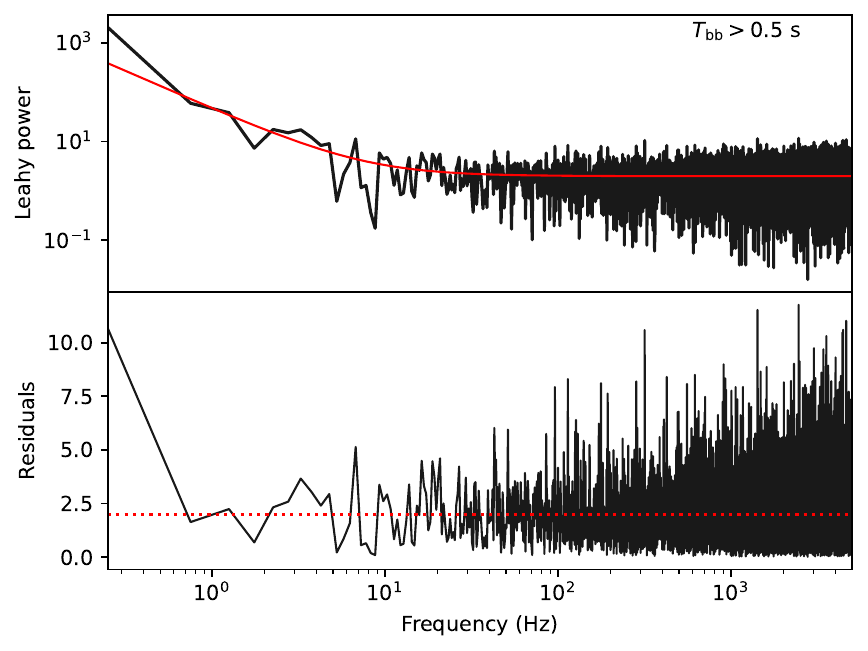}
\end{minipage}
\begin{minipage}[t]{0.49\textwidth}
\centering
\includegraphics[width=\columnwidth]{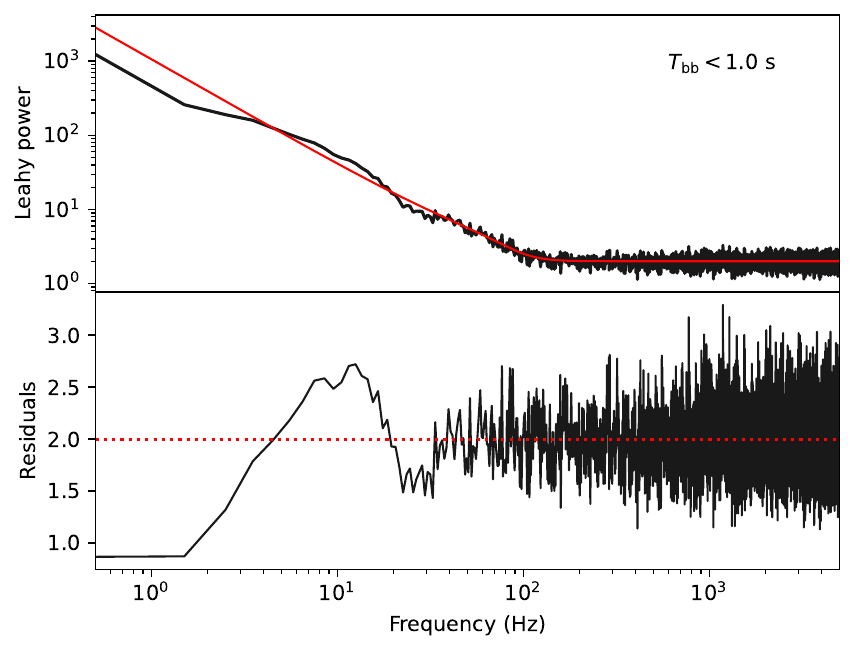}
\end{minipage}
\caption{Averaged periodogram of the bursts with different duration observed by HXMT, the red lines are obtained by fitting with BPLC model. The bottom right panel shows the average power spectrum obtained by choosing 1-second duration for all bursts.}\label{ave_dua_hxmt}
\end{figure*}

\begin{figure}
\centering
\begin{minipage}[t]{0.46\textwidth}
\centering
\includegraphics[width=\columnwidth]{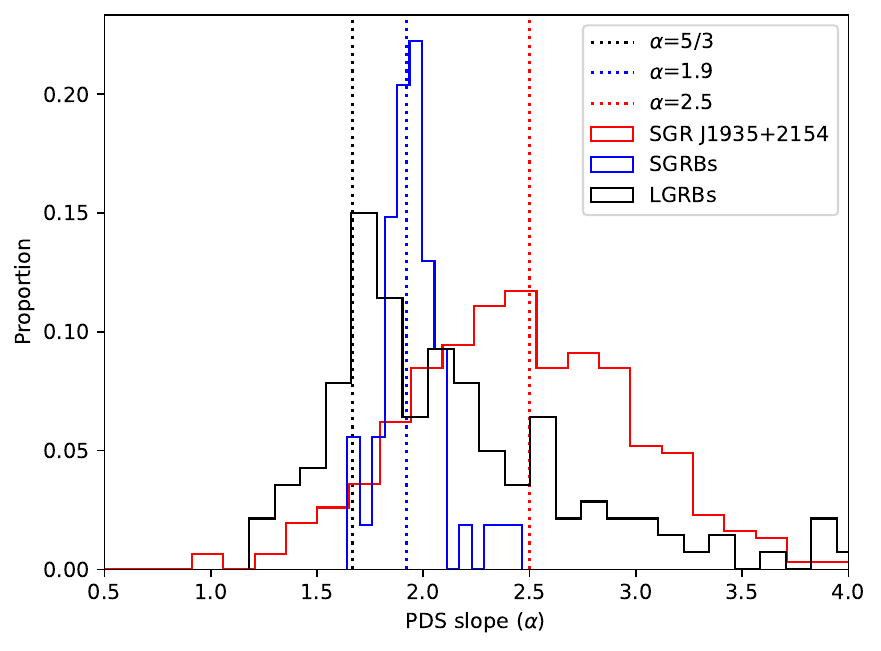}
\end{minipage}
\caption{Comparison of PDS slopes of SGRs and GRBs. Distributions of the PDS slopes ($\alpha$) for bursts from SGR J1935+2154, short and long GRBs. The results of SGRBs and LGRBs are from \citealp{dichiara2013search} and \citealp{guidorzi2016individual}, respectively.}\label{3sources}
\end{figure}

\subsection{Methodology}
\subsubsection{PDS Slopes}
As the power fluctuations of an unbinned PDS follow a $\chi^{2}$ distribution, we first fit the periodogram using a combination of red noise at low frequencies and white noise at high frequencies with a power law plus a constant (PLC) \citep{2010MNRAS.402..307V,huppenkothen2013quasi}, 

\begin{equation}
S_{\rm plc}(f) = \beta f^{-\alpha}+\gamma,
\label{eq:plc}
\end{equation}
where $\alpha$ is the power-law index (i.e. slope), or fitted with a broken power law plus a constant (BPLC) \citep{huppenkothen2013quasi},
\begin{equation}
S_{\rm bplc}(f) = \beta v^{-\alpha_1}(1+(v/\delta)^{(\alpha_2-\alpha_1)})^{-1}+\gamma,
\label{eq:bplc}
\end{equation}
where $\alpha_1$ and $\alpha_2$ are the power-law indices in low and high frequencies (the frequency $\delta$ is the change point), respectively, and $\gamma$ is a constant
that represents the white (Poisson) noise in the
periodogram.

To compare the goodness of fit for the PLC and BPLC models, we use the likelihood ratio test (LRT) in the Bayesian implementation \citep{2010MNRAS.402..307V} and accept the BPLC model when the probability of chance improvement $p$-value  $<$0.05. We perform the above calculations using $Stingray$ package \footnote{\href{https://docs.stingray.science/index.html}{https://docs.stingray.science/index.html}} \citep{huppenkothen2019stingray}.

\subsubsection{Search for QPO signals}
For the periodic signal search, we adopt the steps described in \citealp{huppenkothen2014quasi}, that is, comparing the maximum
residuals $R_{j}^{\rm obj}=2I_{j}^{\rm obs}/S_{j}(\theta)$ of a large number of simulations to the maximum
residual for the best-fit broadband model (i.e. PLC or BPLC) in the observed periodogram. 
To achieve a reasonable coverage of the frequency range, we use 1, 3, 5, 7 and 10 times the original frequency resolution, which can ensure that the searches are sensitive to periodic signals with smaller widths, approximately equal to the unbinned frequency resolution and broader signals for the potential periodic component.

When the probability of no QPO signal $p<1\times10^{-5}$ is in at least one frequency bin for observing the recorded maximum power residual under the assumption of pure noise, we consider it has a potential QPO signal. It is worth noting that the threshold $p<1\times10^{-5}$ is conservative because there are 628 bursts across 5 different frequency resolutions, that is, the modified threshold is $p<0.031$, corresponding to a Gaussian significance of the threshold of only 1.9$\sigma$.

We note that there is another approach to searching for QPOs based on a model selection problem, that is, the broadband noise model is compared to a more complex model combining both the broadband noise model and a Lorentzian (account for the QPO) \citep{huppenkothen2013quasi}. Although this method may better represent the shape of the PDS, but it requires more calculations on intricate model selection criteria; we thus do not utilize the method in this work.

\section{Result}

\subsection{PDS slopes}

As shown in Fig.~\ref{alphas}, the PDS slopes (i.e. power-law index) in the PLC model of the bursts from SGR J1935+2154 are around 2.5 and the distribution ranges observed by the three satellites are consistent (Note only samples with PDS slope errors less than 0.5 are selected). Furthermore, A Kolmogorov–Smirnov (K-S) test comparing the populations observed by GBM and GECAM gives a p-value of 0.04, supporting the null hypothesis that the two populations are drawn from the same underlying distribution. However, the p-value of comparing the populations observed by GBM and HXMT is 0.003, which indicates that they may come from different distributions; however, since the sample of HXMT is small, no significant conclusion can be given.

The left panel in Fig.~\ref{pl_bpl_p} shows the p-value distribution of the likelihood ratios for the PLC and BPLC models, where the proportions of BPLC being the optimal model (i.e. the $p$-value<0.05) observed by HXMT, GECAM and GBM are 0.31, 0.28, and 0.25, respectively, which are consistent. The right panel in Fig.~\ref{pl_bpl_p} shows that the distribution of the break frequency for the optimal model is BPLC and peaks at $\sim$ 60 Hz. 
As discussed in \citealp{guidorzi2016individual}, the PDS reflects the effect of pulses superposition of different time scales in the light curve; if the total variance is mostly dominated by some specific timescale, the PDS will be best fit with BPLC with the timescale corresponding to the break frequency $f_{\rm b}$. Therefore, the distribution of the break frequency peaks at $\sim$ 60 Hz implies that several pulses in some bursts have a duration of $\sim$ 17 ms.

On the other hand, we investigate whether the PDS slopes for the same bursts observed by GECAM and GBM are consistent. As shown in Fig.~\ref{GBM_GECAM}, the results are consistent within the error range. It is worth noting here that the slopes of several bursts are smaller inferred with GECAM data, due to their lower signal-to-noise ratios observed by GECAM. 

\cite{huppenkothen2014quasi} reported a weak anti-correlation (the power-law index is -0.1) between the power-law index of the PDS and duration in bursts from SGR J1550-5418, that is, shorter bursts usually have steeper power-law indices than longer bursts. Therefore, we also investigate the power-law index and duration relationships of bursts from SGR J1935+1935; as shown in the left panel of Fig.~\ref{alpha_T}, they can also be fitted well with a power law $\alpha=2.27_{-0.02}^{+0.02}\times T_{\rm BB}^{-0.06_{-0.01}^{+0.01}}$. Besides, we find a similar anti-correlation power-law relationship between the power-law index and the minimum variation timescale; as shown in the right panel of Fig.~\ref{alpha_T}, they can be fitted well with a power law $\alpha=1.28_{-0.06}^{+0.06}\times \Delta t_{\rm MVT}^{-0.11_{-0.01}^{+0.01}}$, which suggests that the more drastically variable bursts usually have steeper PDS slopes.

\subsection{QPO Searches in Individual Burst}
We find no QPO candidate with significance above 3 $\sigma$; however, this does not exclude that QPO actually exists especially in low frequencies due to the limitations of this PDS-based method used in this work. Since the burst envelope and red noise dominate in the low frequencies, \cite{huppenkothen2013quasi} indicated through simulations that even strong signals below 70 Hz will be very difficult to detect. 
Since \cite{li2022quasi} recently reported a quasi-periodic oscillation (QPO) with 3.4$\sigma$ significance
at $\sim$ 40 Hz in the X-ray burst associated with FRB 200428 from SGR J1935+2154, we carefully examine the bursts with maximum residuals corresponding to frequencies in the range of 30-50 Hz and find that in some bursts (e.g. UTC 2021-09-11T20:13:40.550, UTC 2020-04-27T21:15:36.400, UTC 2019-11-04T09:17:53.480 and UTC 2020-05-16T18:12:52.080) possible QPO at $\sim$ 40 Hz may exist in the data of HXMT, GECAM and GBM. As shown in Figs. \ref{lcps0}, \ref{lcps1}, \ref{lcps2} and \ref{lcps3}, their light curves are all multi-pulses and have a period of about $\sim$ 25 ms, resulting in possible signals at $\sim$ 40 Hz in each PDS. Interestingly, we find features at $\sim$ 80 Hz in bursts UTC 2020-04-27T21:15:36.400 and UTC 2020-05-16T18:12:52.080, which may be harmonics of the 40 Hz QPO. However, none of them are significant enough to be claimed as a QPO; thus we only consider them as candidates.

\subsection{QPO Searches in Average PDS}
Due to the low efficiency of QPO detection at lower frequencies \citep{huppenkothen2013quasi}, we need to hypothesize the possibility that the majority of bursts have QPOs at some frequency (e.g. excite the same star quakes), but cannot be detected in individual bursts due to sensitivity. Therefore, we divide the bursts into five groups based on duration, i.e. $<50$ ms, 50-100 ms, 100-250 ms,250-500 ms and 500-1000 ms. We select the longest duration (i.e. 50 ms, 100 ms, 250 ms, 500 ms and 1000 ms) in each group as the duration of light curves for each burst in the same group; thus each periodogram would have the same number of frequencies to average periodograms.

The average PDSs for the five groups observed by GECAM, GECAM and HXMT are shown in Fig.~\ref{ave_dua},\ref{ave_dua_gecam} and \ref{ave_dua_hxmt}, which are all fitted with BPL. No definitive QPO is detected in the groups; however, some features are detected at about 40 Hz in the longer duration groups 250-500 ms and 500-1000 ms observed by GBM.

\section{Discussion and conclusion}
In this work, we investigate the PDS slopes (including the distribution, and the dependence on the duration and the minimum variation timescale) and QPO searches (including individual bursts and average PDS) for hundreds of bursts from SGR J1935+2154 observed by HXMT, GECAM and GBM.

If a p-value of 0.05 is taken as the threshold, about 30\% of the bursts can be better fitted with the BPLC model and the others should be preferred with the PLC model. The PDS slope distribution for PLC model peaks at $\sim$ 2.5, which are consistent with other magnetars, such as SGR J0501+4516 \citep{huppenkothen2013quasi} and SGR J1550-5418 \citep{huppenkothen2014quasi}. The distributions obtained from the three satellites are not significantly different.
Besides, we also report a power law relationship $\alpha=2.27_{-0.02}^{+0.02}\times T_{\rm BB}^{-0.06_{-0.01}^{+0.01}}$ between the power-law index of PDS and the burst duration, and the power-law index is roughly consistent with the result in J1550-5418 \citep{huppenkothen2014quasi}. Interestingly, we discover a power-law relationship $\alpha=1.28_{-0.06}^{+0.06}\times \Delta t_{\rm MVT}^{-0.11_{-0.01}^{+0.01}}$ between the power-law index of PDS and the minimum variation timescale, which suggests that the more drastically variable bursts usually have a steeper PDS slope. On the other hand, the distribution of the break frequency for the optimal model is BPLC and peaks at $\sim$ 60 Hz, which implies that the pulses in some bursts have a duration of $\sim$ 17 ms.

We also compare the PDS slope distributions of SGR J1935+2154 and LGRBs, SGRBs, where the results of SGRBs and LGRBs are from \citealp{dichiara2013search} and \citealp{guidorzi2016individual}. As shown in Fig.~\ref{3sources}, their distributions are different, and the peaks are around 2.5, 5/3, and 1.9, respectively. Based on the anti-correlation between the PDS slope and the minimum variation timescale, we speculate that the reason for the steeper slope of SGR J1935+2154 is due to its smaller minimum variation timescale (the median value of the distribution is $\sim$ 2 ms) \citep{xiao2023minimum}, but the median values for the SGRB and LGRB are 40 ms and 480 ms \citep{golkhou2015energy}, respectively.

Although QPOs with sufficient significance ($>3\sigma$) are not found in the bursts from SGR J1935+2154, we can not exclude that QPOs actually exist especially in low frequencies, due to the detection difficulties of the PDS-based method \citep{huppenkothen2014quasi} used in this work. In fact, The most straightforward way is to model the
shape of the light curve directly, and then compare the observed periodogram to the periodograms of the simulated light curves (e.g. \citealp{fox2001search}). However, due to the absence of knowledge about the emission processes in magnetar bursts (i.e. the light curves of magnetar bursts cannot be well modeled), we adopt the conservative
method that models the burst light curve in PDS as a pure
red noise process.

We detect some features around 40 Hz in the PDS of multiple bursts, the frequency of which is consistent with that in the burst in association with FRB 200428 \citep{li2022quasi}. Furthermore, the features also appear in the average power density spectra of bursts with durations from 250 to 1000 ms. 
These imply that the 40 Hz QPO in the burst associated with FRB 200428 may be not peculiar. In addition, the frequency of 40 Hz also lies in the range of 18-155 Hz of the QPO of the giant flares from other magnetars such as SGR 1900+14 \citep{strohmayer2005discovery} and SGR 1806-20 \citep{watts2006detection}. According to previous studies, the 40 Hz QPO can be explained as torsional Alfven modes of the highly magnetized fluid core \citep{2001ApJ...561..980T} or torsional shear modes of the neutron star crust \citep{2005ApJ...634L.153P}. Therefore, the detected QPOs are potentially powerful to constrain the interior structures of magnetars \citep{2007PhR...442..109L}.


\section*{Acknowledgments}
We acknowledge the public data from {\it Fermi}/GBM. This work is supported by the National Key R\&D Program of China (2022YFF0711404, 2021YFA0718500), the Scientific Research Project of the Guizhou Provincial Education (Nos. KY[2022]123, KY[2022]132, KY[2022]137, KY[2021]303, KY[2020]003 and KY[2023]059), and partially supported by International Partnership Program of Chinese Academy of Sciences (Grant No. 113111KYSB20190020). 
The authors also thank supports from 
the Strategic Priority Research Program on Space Science, the Chinese Academy of Sciences (Grant No.
XDA15010100, 
XDA15360100, XDA15360102, 
XDA15360300, 
XDA15052700), 
the National Natural Science Foundation of China (Projects: 12273042, 12061131007, Grant No. 12173038, Nos. 12273008 and 12103013), the Foundation of Education Bureau of Guizhou Province, China (Grant No. KY (2020) 003), Guizhou Provincial Science and Technology Foundation (Nos. ZK[2022]304, ZK[2022]322, [2021]023, [2023]024), the National SKA Program of China (Nos. 2022SKA0130100, 2022SKA0130104), the Major Science and Technology Program of Xinjiang Uygur Autonomous Region (No.2022A03013-4), the Joint Research Fund in Astronomy (Grant Nos. U1931101) under cooperative agreement between the National Natural Science Foundation of China (NSFC) and Chinese Academy of Sciences (CAS). S. Xiao is grateful to W. Xiao, G. Q. Wang, J. H. Li and Q. Q. Xiao for their useful comments. 

\newpage

\bibliography{main}

\begin{thebibliography}{}
\expandafter\ifx\csname natexlab\endcsname\relax\def\natexlab#1{#1}\fi
\providecommand{\url}[1]{\href{#1}{#1}}
\providecommand{\dodoi}[1]{doi:~\href{http://doi.org/#1}{\nolinkurl{#1}}}
\providecommand{\doeprint}[1]{\href{http://ascl.net/#1}{\nolinkurl{http://ascl.net/#1}}}
\providecommand{\doarXiv}[1]{\href{https://arxiv.org/abs/#1}{\nolinkurl{https://arxiv.org/abs/#1}}}

\bibitem[{Castro-Tirado {et~al.}(2021)Castro-Tirado, {\O}stgaard,
  G{\"o}ǧ{\"u}{\c{s}}, S{\'a}nchez-Gil, Pascual-Granado, Reglero, Mezentsev,
  Gabler, Marisaldi, Neubert, {et~al.}}]{castro2021very}
Castro-Tirado, A.~J., {\O}stgaard, N., G{\"o}ǧ{\"u}{\c{s}}, E., {et~al.} 2021,
  Nature, 600, 621

\bibitem[{Dichiara {et~al.}(2013)Dichiara, Guidorzi, Frontera, \&
  Amati}]{dichiara2013search}
Dichiara, S., Guidorzi, C., Frontera, F., \& Amati, L. 2013, The Astrophysical
  Journal, 777, 132

\bibitem[{Fox {et~al.}(2001)Fox, Lewin, Rutledge, Morgan, Guerriero, Bildsten,
  van~der Klis, Van~Paradijs, Moore, Dotani, {et~al.}}]{fox2001search}
Fox, D., Lewin, W., Rutledge, R., {et~al.} 2001, Monthly Notices of the Royal
  Astronomical Society, 321, 776

\bibitem[{{Gao} {et~al.}(2012){Gao}, {Zhang}, \& {Zhang}}]{2012ApJ...748..134G}
{Gao}, H., {Zhang}, B.-B., \& {Zhang}, B. 2012, \apj, 748, 134,
  \dodoi{10.1088/0004-637X/748/2/134}

\bibitem[{Golkhou {et~al.}(2015)Golkhou, Butler, \&
  Littlejohns}]{golkhou2015energy}
Golkhou, V.~Z., Butler, N.~R., \& Littlejohns, O.~M. 2015, The Astrophysical
  Journal, 811, 93

\bibitem[{Guidorzi {et~al.}(2016)Guidorzi, Dichiara, \&
  Amati}]{guidorzi2016individual}
Guidorzi, C., Dichiara, S., \& Amati, L. 2016, Astronomy \& Astrophysics, 589,
  A98

\bibitem[{Huppenkothen {et~al.}(2014)Huppenkothen, Heil, Watts, \&
  G{\"o}{\u{g}}{\"u}{\c{s}}}]{huppenkothen2014quasi}
Huppenkothen, D., Heil, L., Watts, A., \& G{\"o}{\u{g}}{\"u}{\c{s}}, E. 2014,
  The Astrophysical Journal, 795, 114

\bibitem[{{Huppenkothen} {et~al.}(2014{\natexlab{a}}){Huppenkothen}, {Watts},
  \& {Levin}}]{2014ApJ...793..129H}
{Huppenkothen}, D., {Watts}, A.~L., \& {Levin}, Y. 2014{\natexlab{a}}, \apj,
  793, 129, \dodoi{10.1088/0004-637X/793/2/129}

\bibitem[{Huppenkothen {et~al.}(2013)Huppenkothen, Watts, Uttley, Van~der
  Horst, Van~der Klis, Kouveliotou, G{\"o}{\u{g}}{\"u}{\c{s}}, Granot, Vaughan,
  \& Finger}]{huppenkothen2013quasi}
Huppenkothen, D., Watts, A.~L., Uttley, P., {et~al.} 2013, The Astrophysical
  Journal, 768, 87

\bibitem[{{Huppenkothen} {et~al.}(2014{\natexlab{b}}){Huppenkothen},
  {D'Angelo}, {Watts}, {Heil}, {van der Klis}, {van der Horst}, {Kouveliotou},
  {Baring}, {G{\"o}{\u{g}}{\"u}{\c{s}}}, {Granot}, {Kaneko}, {Lin}, {von
  Kienlin}, \& {Younes}}]{2014ApJ...787..128H}
{Huppenkothen}, D., {D'Angelo}, C., {Watts}, A.~L., {et~al.}
  2014{\natexlab{b}}, \apj, 787, 128, \dodoi{10.1088/0004-637X/787/2/128}

\bibitem[{Huppenkothen {et~al.}(2019)Huppenkothen, Bachetti, Stevens, Migliari,
  Balm, Hammad, Khan, Mishra, Rashid, Sharma,
  {et~al.}}]{huppenkothen2019stingray}
Huppenkothen, D., Bachetti, M., Stevens, A.~L., {et~al.} 2019, The
  Astrophysical Journal, 881, 39

\bibitem[{Israel {et~al.}(2005)Israel, Belloni, Stella, Rephaeli, Gruber,
  Casella, Dall’Osso, Rea, Persic, \& Rothschild}]{israel2005discovery}
Israel, G., Belloni, T., Stella, L., {et~al.} 2005, The Astrophysical Journal,
  628, L53

\bibitem[{{Lattimer} \& {Prakash}(2007)}]{2007PhR...442..109L}
{Lattimer}, J.~M., \& {Prakash}, M. 2007, \physrep, 442, 109,
  \dodoi{10.1016/j.physrep.2007.02.003}

\bibitem[{Li {et~al.}(2022)Li, Ge, Lin, Zhang, Song, Cao, Zhang, Lu, Xu, Xiong,
  {et~al.}}]{li2022quasi}
Li, X., Ge, M., Lin, L., {et~al.} 2022, The Astrophysical Journal, 931, 56

\bibitem[{Liu {et~al.}(2020)Liu, Zhang, Li, Lu, Chang, Li, Zhang, Jin, Yu,
  Zhang, {et~al.}}]{liu2020high}
Liu, C., Zhang, Y., Li, X., {et~al.} 2020, SCIENCE CHINA Physics, Mechanics \&
  Astronomy, 63, 1

\bibitem[{Liu {et~al.}(2021)Liu, Gong, Li, Wen, An, Cai, Chang, Chen, Chen, Du,
  {et~al.}}]{liu2021sipm}
Liu, Y., Gong, K., Li, X., {et~al.} 2021, arXiv preprint arXiv:2112.04786

\bibitem[{Meegan {et~al.}(2009)Meegan, Lichti, Bhat, Bissaldi, Briggs,
  Connaughton, Diehl, Fishman, Greiner, Hoover, {et~al.}}]{meegan2009fermi}
Meegan, C., Lichti, G., Bhat, P., {et~al.} 2009, The Astrophysical Journal,
  702, 791

\bibitem[{{Miller} {et~al.}(2019){Miller}, {Chirenti}, \&
  {Strohmayer}}]{2019ApJ...871...95M}
{Miller}, M.~C., {Chirenti}, C., \& {Strohmayer}, T.~E. 2019, \apj, 871, 95,
  \dodoi{10.3847/1538-4357/aaf5ce}

\bibitem[{{Piro}(2005)}]{2005ApJ...634L.153P}
{Piro}, A.~L. 2005, \apjl, 634, L153, \dodoi{10.1086/499049}

\bibitem[{Pumpe {et~al.}(2018)Pumpe, Gabler, Steininger, \&
  En{\ss}lin}]{pumpe2018search}
Pumpe, D., Gabler, M., Steininger, T., \& En{\ss}lin, T.~A. 2018, Astronomy \&
  Astrophysics, 610, A61

\bibitem[{Roberts {et~al.}(2023)Roberts, Baring, Huppenkothen, Gogus, Kaneko,
  Kouveliotou, Lin, van~der Horst, \& Younes}]{roberts2023quasi}
Roberts, O.~J., Baring, M.~G., Huppenkothen, D., {et~al.} 2023, arXiv preprint
  arXiv:2306.08130

\bibitem[{Scargle {et~al.}(2013)Scargle, Norris, Jackson, \&
  Chiang}]{scargle2013studies}
Scargle, J.~D., Norris, J.~P., Jackson, B., \& Chiang, J. 2013, The
  Astrophysical Journal, 764, 167

\bibitem[{{Shen} \& {Song}(2003)}]{2003PASJ...55..345S}
{Shen}, R.-F., \& {Song}, L.-M. 2003, \pasj, 55, 345,
  \dodoi{10.1093/pasj/55.2.345}

\bibitem[{Strohmayer \& Watts(2005)}]{strohmayer2005discovery}
Strohmayer, T.~E., \& Watts, A.~L. 2005, The Astrophysical Journal, 632, L111

\bibitem[{{Strohmayer} \& {Watts}(2006)}]{2006ApJ...653..593S}
{Strohmayer}, T.~E., \& {Watts}, A.~L. 2006, \apj, 653, 593,
  \dodoi{10.1086/508703}

\bibitem[{{Thompson} \& {Duncan}(2001)}]{2001ApJ...561..980T}
{Thompson}, C., \& {Duncan}, R.~C. 2001, \apj, 561, 980, \dodoi{10.1086/323256}

\bibitem[{{Vaughan}(2010)}]{2010MNRAS.402..307V}
{Vaughan}, S. 2010, \mnras, 402, 307, \dodoi{10.1111/j.1365-2966.2009.15868.x}

\bibitem[{Vetere {et~al.}(2006)Vetere, Massaro, Costa, Soffitta, \&
  Ventura}]{vetere2006slow}
Vetere, L., Massaro, E., Costa, E., Soffitta, P., \& Ventura, G. 2006,
  Astronomy \& Astrophysics, 447, 499

\bibitem[{Watts \& Strohmayer(2006)}]{watts2006detection}
Watts, A.~L., \& Strohmayer, T.~E. 2006, The Astrophysical Journal, 637, L117

\bibitem[{Xiao {et~al.}(2020)Xiao, Xiong, Liu, Li, Zhang, Ge, Cai, Yi, Zhu,
  Chen, {et~al.}}]{xiao2020deadtime}
Xiao, S., Xiong, S., Liu, C., {et~al.} 2020, Journal of High Energy
  Astrophysics, 26, 58

\bibitem[{Xiao {et~al.}(2022{\natexlab{a}})Xiao, Zhang, Zhu, Xiong, Gao, Xu,
  Zhang, Peng, Li, Zhang, {et~al.}}]{xiao2022quasi}
Xiao, S., Zhang, Y.-Q., Zhu, Z.-P., {et~al.} 2022{\natexlab{a}}, arXiv preprint
  arXiv:2205.02186

\bibitem[{Xiao {et~al.}(2022{\natexlab{b}})Xiao, Peng, Zhang, Xiong, Li, Tuo,
  Gao, Wang, Xue, Zheng, {et~al.}}]{xiao2022search}
Xiao, S., Peng, W.-X., Zhang, S.-N., {et~al.} 2022{\natexlab{b}}, The
  Astrophysical Journal, 941, 166

\bibitem[{Xiao {et~al.}(2022{\natexlab{c}})Xiao, Liu, Peng, An, Xiong, Tuo,
  Gong, Zhang, Zhang, Zheng, {et~al.}}]{xiao2022ground}
Xiao, S., Liu, Y., Peng, W., {et~al.} 2022{\natexlab{c}}, Monthly Notices of
  the Royal Astronomical Society, 511, 964

\bibitem[{{Xiao} {et~al.}(2023{\natexlab{a}}){Xiao}, {Tuo}, {Zhang}, {Xiong},
  {Lin}, {Zhang}, {Wang}, {Xue}, {Cai}, {Gao}, {Li}, {Li}, {Zheng}, {Liu},
  {Wang}, {Wang}, {Peng}, {Liu}, {Li}, {Wen}, {An}, {Song}, {Zheng}, {Zhang},
  {Dong}, {Xie}, {Feng}, {Ma}, {Wang}, {Luo}, {Dang}, {Shang}, {Zhi}, \&
  {Li}}]{xiao2023discovery}
{Xiao}, S., {Tuo}, Y.-L., {Zhang}, S.-N., {et~al.} 2023{\natexlab{a}}, \mnras,
  521, 5308, \dodoi{10.1093/mnras/stad885}

\bibitem[{{Xiao} {et~al.}(2023{\natexlab{b}}){Xiao}, Yang, Luo, Xiong, Qu,
  Zhang, Xue, Li, Tuo, Dong, Zhao, Dang, Shang, Ma, Cai, Wang, Wang, Li, Yi,
  Zhang, Ge, Zheng, Song, Peng, Wen, Li, An, Xu, Wang, Zheng, Zhang, Liu,
  Zhang, Xie, Feng, Wang, \& Zhi}]{xiao2023minimum}
{Xiao}, S., Yang, J.-J., Luo, X.-H., {et~al.} 2023{\natexlab{b}}.
\newblock \doarXiv{2307.07079}

\end{thebibliography}

\end{CJK}
\end{document}